\documentclass[a4paper,12pt]{article}
\usepackage{amssymb}
\usepackage{amsfonts}
\usepackage{afterpage}
\usepackage{dsfont}
\usepackage{bbm}
\usepackage{booktabs}
\usepackage{amsmath}
\usepackage{graphicx}
\usepackage{amsthm}
\usepackage{url}
\usepackage{pdflscape}
\usepackage{array}
\usepackage{arydshln}
\usepackage{verbatim}
\usepackage{pifont}
\usepackage{eurosym}
\usepackage[authoryear]{natbib}
\usepackage{tablefootnote}
\usepackage{multirow}							% For Tables
\usepackage{arydshln}							% For Dash lines in Tables
\usepackage{adjustbox}
\usepackage{caption}
\usepackage{dsfont}
\usepackage{natbib}
\usepackage[colorlinks=true, linkcolor=blue, citecolor=blue, urlcolor=blue, linktocpage=true]{hyperref} 

\newcommand*{\transpose}{^{\mkern-1.5mu\mathsf{T}}}

\DeclareMathOperator*{\argmin}{arg\,min}
\makeatletter
\def\@biblabel#1{}
\makeatother

\usepackage{color}
\usepackage{verbatim}
\definecolor{kit-blue}{RGB}{70,100,170}

\def\boxit#1{\vbox{\hrule\hbox{\vrule\kern6pt\vbox{\kern6pt#1\kern6pt}\kern6pt\vrule}\hrule}}

\newcolumntype{L}[1]{>{\raggedright\let\newline\\\arraybackslash\hspace{0pt}}m{#1}}

\begin{document}

\title{\vspace{-2.0cm} \Large How have German University Tuition Fees Affected Enrollment Rates: Robust Model Selection and Design-based Inference in High-Dimensions} 

\author{Konstantin G\"{o}rgen \thanks{Konstantin G\"{o}rgen, email: konstantin.goergen@kit.edu; phone: +49/721/608/43793}\\ \normalsize Karlsruhe Institute of Technology, Germany \and Melanie Schienle \thanks{Melanie Schienle, email: melanie.schienle@kit.edu, phone: +49/721/60847535, both: Department of Economics (ECON), Karlsruhe Institute of Technology, Bl\"{u}cherstr.17, 76185 Karlsruhe, Germany}  \\ \normalsize {Karlsruhe Institute of Technology, Germany}}
\date{\normalsize
	\today
}
\maketitle

%%%%%%%%%%%%%%%%%%%%%%%%%%%%%%%%%%%%%%%%%%%%%%%%%%%%%%%%%%%%%%%%%%%%%%%%%%%%%%%%%%%%%%%%%%%%%%%%%%%%%%%%%
%%%%%%%%%%%%%%%%%%%%%%%%%%%%%%%%%%%%%%%%%%%%%%%%%%%%%%%%%%%%%%%%%%%%%%%%%%%%%%%%%%%%%%%%%%%%%%%%%%%%%%%%%

%\vspace{-0.5cm}
\thispagestyle{empty}
\begin{abstract}
\noindent We use official data for all 16 federal German states to study the causal effect of a flat 1000 Euro state-dependent university tuition fee on the enrollment behavior of students during the years 2006-2014. In particular, we show how the variation in the introduction scheme across states and times can be exploited to identify the federal average causal effect of tuition fees by controlling for a large amount of potentially influencing attributes for state heterogeneity. We suggest a stability post-double selection methodology to robustly determine the causal effect across types in the transparently modeled unknown response components. The proposed stability resampling scheme in the two LASSO selection steps efficiently mitigates the risk of model underspecification and thus biased effects when the tuition fee policy decision also depends on relevant variables for the state enrollment rates. Correct inference for the full cross-section state population in the sample requires adequate design- rather than sampling-based standard errors.
With the data-driven model selection and explicit control for spatial cross-effects we detect that tuition fees induce substantial migration effects where the mobility occurs both from fee but also from non-fee states suggesting also a general movement for quality. Overall, we find a significant negative impact of up to 4.5 percentage points of fees on student enrollment. This is in contrast to plain one-step LASSO or previous empirical studies with full fixed effects linear panel regressions which generally underestimate the size and get an only insignificant effect.

	\vspace{0,5cm}
	
	
	%\vspace{0,5cm}
	
	 \vspace{0.5in}
\end{abstract}

%%%%%%%%%%%%%%%%%%%%%%%%%%%%%%%%%%%%%%%%%%%%%%%%%%%%%%%%%%%%%%%%%%%%%%%%%%%%%%%%%%%%%%%%%%%%%%%%%%%%%%%%%
%%%%%%%%%%%%%%%%%%%%%%%%%%%%%%%%%%%%%%%%%%%%%%%%%%%%%%%%%%%%%%%%%%%%%%%%%%%%%%%%%%%%%%%%%%%%%%%%%%%%%%%%%

\pagebreak

\section{Introduction}

In this paper, we study the causal effect of the introduction of a flat state-dependent tuition fee on university student enrollment behavior using official data for all 16 federal German states. In particular, we show how the variation in the introduction scheme across states and time can be used to identify the federal average causal effect of tuition fees by controlling for a large amount of potentially influencing attributes for state heterogeneity. In Germany, in contrast to other countries, the maximum fee amount was generally limited to {1000} Euros per year and fees were only present in parts of the country from 2006-2014\footnote{We always observe year $t$ at the beginning of the winter term in October of $t$.}. Moreover, the implementation and timing of the fees, both, were no exogenous shock but evidence driven policy decisions on the federal state level (``Bundesl\"{a}nder'', denoted as states in the following) and thus varied considerably among states. At the same time, however, major policy changes in different federal states also significantly impacted the cohort size of prospective university students.\footnote{This comprises a decrease for the required compulsory years to high school graduation from nine to eight years of which the introduction varied on the state level, and the general German-wide abolishment of the 9 month compulsory military service for men in the age of 17-23.} This spatial time delay in the implementation of both tuition fees and different federal reforms induced substantial migration effects which potentially impacted state-level student enrollment on top of the many standard socio-economic state characteristics.

We thus suggest a stability post-double selection methodology (cp.\citet{Belloni2014b}) to robustly determine the causal effect in such a high-dimensional setting with many potentially influential controls and few observations with measurement problems. With a robust subsampling-augmented Lasso procedure (cp. \citet{Meinshausen2010}), we adaptively select the relevant controls not only in the outcome equation, but also crucially augment this set with the Lasso selection choices in an auxiliary propensity score equation. Given the strong correlation of the tuition fee decision and the control variables, this double-selection type strategy ensures that underspecification and resulting biased estimates are not an issue. Overall, with these tailored data-driven techniques, we detect a significant negative effect of tuition fees inducing an up to $4.5$ percentage point (pp) reduction in enrollment rates. Since the exact enrollment rate is hard to measure, we show the stability of our results over a large grid of values and we employ design-based standard errors which reflect that the full population of states enters the estimation. While spatial cross-effects have been ignored in the previous literature on German tuition fees (see e.g. \citet{Dwenger2012, Bruckmeier2014,Mitze2015})), we identify them as important drivers for enrollment rates by the Lasso, besides state specific factors such as the student-to-researcher ratio. We explicitly show that without Lasso pre-selection of variables, the signal to noise ratio of the problem is too low for detecting the correct magnitude of the effect. Generally, these insights and our methodological solution are highly relevant for all cases of policy evaluation, where implementation occurs in a spatially time-delayed manner, as for example environmental policies that target global warming or financial regulations in different countries. In addition, we believe that our empirical findings cannot only contribute to the active ongoing discussions on reintroducing tuition fees in Germany, but might also be of independent interest for other countries such as the United Kingdom, where fees are on the rise.\\

For the analysis we study the years 2005-2014 and all 16 federal states in Germany. We include a comprehensive set of 18 covariates, covering all potentially important controls of the national and international literature on tuition fee effects (e.g. \citet{Dynarski2003, Kane1994} and \citet{Baier2011, Dwenger2012,Bruckmeier2014,Mitze2015}). The variables are collected from different sources, but public data on student enrollment behavior is only available on the state level and not on a university level, which is due to strict German data protection laws.\footnote{Note that across states and universities, individual or household panel data from common sources such as e.g. the German SOEP is insufficient, incomplete and very unbalanced and cannot be employed for a general analysis. Please see Appendix \ref{Sec:Appendix_SOEP_1} for details.} In addition to standard economic, social and educational factors from the literature on student enrollment rates, we also include specific effects for Germany which play a major role in the considered period. Particularly, policy changes such as the abolishment of mandatory military service or the heterogeneous introduction of a one-year reduced secondary education ("G8") in different states are key policies. Moreover, in addition to the above standard list of controls, we construct spatial variables that capture state cross-effects in the policy decisions for or against fees as the proportion of students migrating to each state from states with and without tuition fees based on their proximity. These are crucial to control for migration effects due to heterogeneous implementation and time delay of policies across states that could otherwise bias the estimated effect of tuition fees.
We work with relative enrollment rates instead of absolute numbers as the dependent variable to ensure compatibility of effects across federal states of different population sizes. For correct ratios, however, we require the population size of all high school graduates affected by the introduction of tuition fees in a specific state. This quantity is hard to measure and thus prone to measurement errors as it consists not only of recent and less recent high school graduates from this specific state, but also of parts of cohorts from other states and abroad from where students migrate to study. We transparently treat this measurement ambiguity and thus provide results that are robust in this respect.
Overall, the limitation to only state-level data results in a relatively small number of available observations where single observations could gain substantial influence on the overall result. Thus in total, we face a situation of many potentially influential but correlated covariates and relatively few observations with possible outliers due to data quality problems.\\

We tackle these challenges with a tailored subsampling-augmented variable selection technique in a fixed effects panel regression with many controls. The Lasso type double selection is key for avoiding underspecification in the outcome equation since the tuition fee policy treatment decision is strongly correlated with observed controls (see \citet{Belloni2014a,Belloni2014b}). In this, the data-driven choice of covariates from the auxiliary propensity score equation is used to complement the Lasso-determined active set of relevant regressors in the outcome equation allowing for unbiased estimation of the causal effect. For both selection steps, we propose a subsampling based stability selection (see \citet{Meinshausen2010}) in order to mitigate correlation effects among covariates and measurement issues in the available small set of observations. In such cases, pure Lasso might have difficulties in correctly predicting the influence of each variable, which can lead to the choice of too many variables. We illustrate in a thorough simulation study for such challenging situations, that the suggested stability selection substantially improves on the robustness of the selection results in finite samples leading to augmented post-selection estimation results. Given the scarcity of the available public data and the complexity of the setting, the estimated specification in both the outcome and the auxiliary equation is set as linear which allows for the direct identification of the causal effect. Moreover, for correct inference, adequate design- rather than sampling-based standard errors can be obtained (\citet{Abadie2020a}). These account for the fact that the full cross-section population of states is observed and employed for estimation. Thus the uncertainty in the determination of the causal effect does not result from sampling but from unobserved counterfactuals (see also \citet{Manski_2018}).\\

Our set-up corresponds to the high-dimensional machine learning driven causal literature (see \citet{Belloni2014b, Belloni2016b} and \citet{Athey2017} for a survey as well as applications in labor \citet{angrist2019}) for the estimation of average treatment effects. For standard panel settings with a common treatment timing and sufficient time observations there also exist extensions, e.g. by \citet{Athey2016, cher2018, Athey2018} with applications e.g. in labor \citet{Lechner2019}. Generally, our setting is also deeply routed in the standard low-dimensional treatment effects literature retrieving the (average) causal effect of a policy or treatment in a potential outcomes framework (see e.g. \citet{Rubin1974}, \citet{Rubin1977}). In our case, however, standard methods as e.g. simple difference-in-differences \citep{Card1994, Ashenfelter1985}, low-dimensional propensity score or matching techniques (see e.g. \citet{Rosenbaum1983} or for an overview on nonparametric, nonlinear methods \citet{ImbensGW2004}) or simple one-step LASSO variants thereof cannot adapt to the short available time span and few states in order to detect the tuition fee effect.

Up to our knowledge, the literature on student enrollment behavior generally works with only small sets of covariates on which there is no consensus and often subset selection is only ad-hoc or based on heuristics. Therefore, we propose a data-driven statistical procedure in order to empirically identify relevant factors. Nevertheless, there are analyses on effects of tuition fees in various countries that mostly find significant effects only for certain subgroups of the population. \citet{Kane1994}, \citet{Noorbakhsh2002} and \citet{Mcpherson1991}  find negative effects of tuition fees\footnote{In the study of \citet{Mcpherson1991}, the authors find that the net costs (tuition fees minus student aid) have a negative impact, which is an even stronger argument.} for low-income groups or groups with African-American ethnicity for the US.  More generally, \citet{Neill2009} finds that an increase in tuition fees reduces enrollments significantly for the Canadian system. With the availability of individual data in the presence of much higher fees, but also an established scholarship system, US and Canadian studies can identify effects of tuition fees on enrollment that range between $-2.5$pp and $-6.8$pp. For countries where the situation is more comparable to the German system, and the particular case of Germany, previous studies generally cannot to detect significant effects of tuition fees on enrollment rates (see e.g. for Germany \cite{Baier2011,Hubner2012,Dwenger2012,Bruckmeier2014,Mitze2015}, but also \citet{Huijsman1986} for the Netherlands  and \citet{Denny2014} for Ireland). This seems to be caused by the small number of included covariates, while missing out on the key ones according to our statistical selection technique. Variables possibly correlated with the tuition fee decision are mostly ignored, as well as state cross effects through differences in timing, which we show both to be relevant. Moreover, we cover the comprehensive list of all German tuition fee periods and states, which helps to increase precision of estimated effects in contrast to previous studied, who focused only on subperiods, specific states or subgroups. With mostly insignificant effects between $-0.4$pp and $-2.69$pp, the previous German studies seem to systematically underestimate the true impact of fees.\\

The remainder of the paper is structured as follows. A description of the data set and variables is presented in Section \ref{Sec:Data}. It also contains the transparent construction of (a set of) response variables from the limited available information. Section \ref{Section:model} introduces the linear panel model and the Lasso-type selection methods featuring the stability double selection. In Section \ref{Sec:Simulation}, a Monte Carlo simulation shows the advantages of these methods with different distortions in a controlled environment. After discussing the main results of our empirical study in Section \ref{Sec:Empirical}, we conclude in Section \ref{Sec:Conclusions}.

\section{Data} \label{Sec:Data}
\begin{figure}[!htb]
	\centering
	\includegraphics[width=0.95\textwidth]{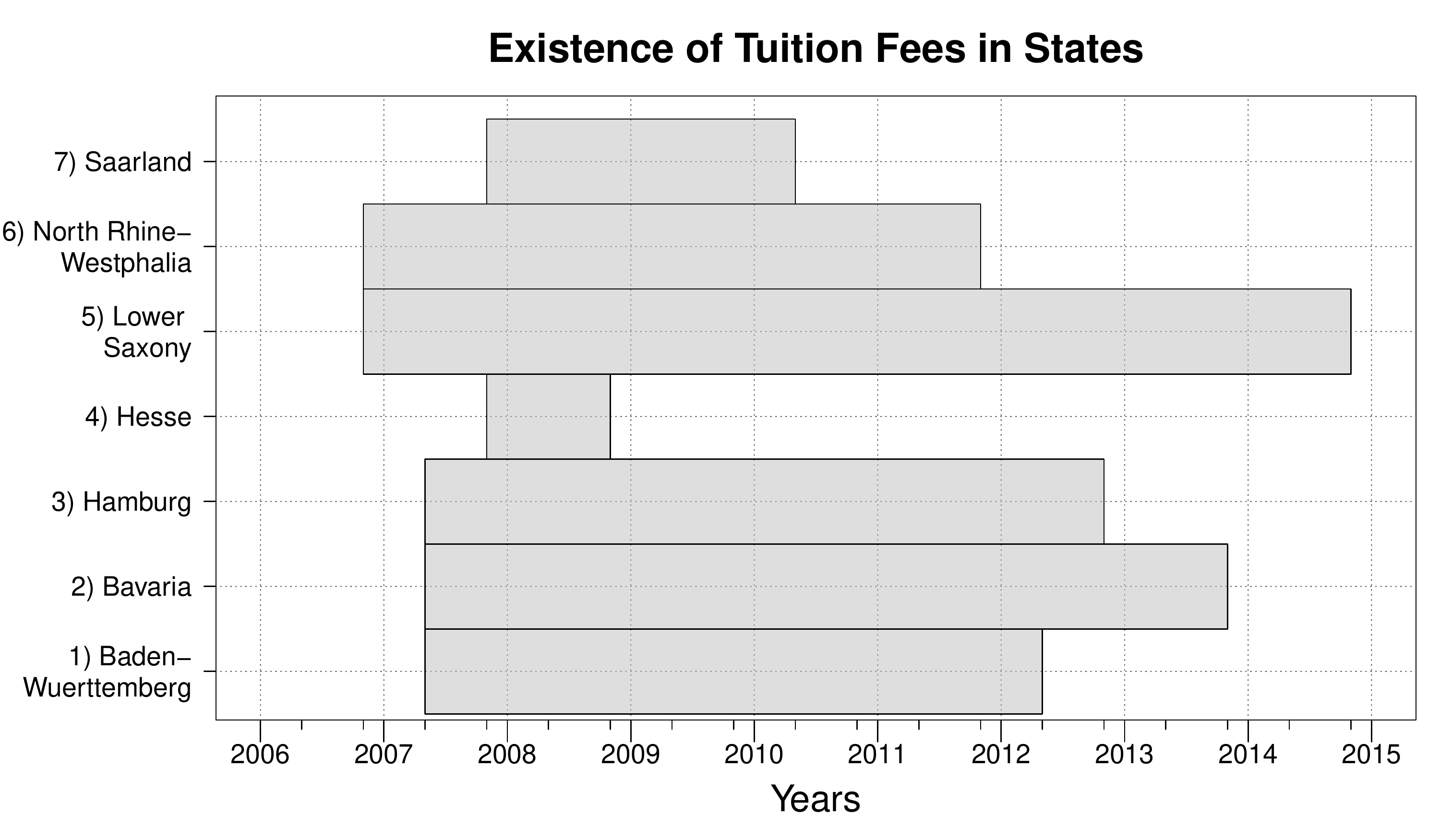}
	\caption{Overview of the timing of tuition fees in German states (presence in gray). The winter term (starting October) and summer term (starting April) are indicated with small ticks. States not listed had no tuition fees at all.}
	\label{fig:tuition_time}
\end{figure}

We construct a panel from publicly available data on enrollment numbers and socio-economic and university-related covariates for the 16 German states ($n=16$) in the years 2005 to 2014 ($T=10$). We use a widespread set of potential controls for determining the effect of tuition fees, which only existed in the years 2006-2014 in at least one state (see Figure ~\ref{fig:tuition_time} for an overview of the timing of fees in different states). The years 2005 and 2014 serve as a base for comparison before and after the introduction and complete abolishment of tuition fees\footnote{As the only state, Lower Saxony abolished Tuition Fees only by the end of the summer term 2014, which is why we still use 2014 as a base for total abolishment of fees.}. Note that we are limited to state level aggregated data, since available individual or household type data from common sources such as e.g. the German Socio-Economic Panel (SOEP) is highly incomplete and very unevenly distributed across states and universities and thus cannot be employed for a general analysis on the effects of tuition fees. Please see Appendix \ref{Sec:Appendix_SOEP_1} for details.

As the response variable we use the enrollment rate $y_{i,t}$ of high school graduates into university in state $i$ at the winter term (WT) of year $t$ to $t+1$ (denoted as $t/t+1$).\footnote{The academic year starts with the winter semester usually beginning in September or October of year $t$ and ending in February of year $t+1$. We use data from public institutions, which account for the majority (more than 90\%) of higher educational institutions in Germany. As higher educational institutions, we denote general university type institutions comprising universities, specialized technical, arts and music universities but also universities of applied sciences (Fachhochschule) and cooperative state universities (Duale Hochschule).} As the population size among German states varies substantially, relative enrollment rates $y_{i,t}$ ensure comparability of results across states, in contrast to the absolute number of new enrollments (from anywhere) $\mathit{NE}_{i,t}$ in state $i$ at WT of year $t/t+1$. The percentage $y_{i,t}$ is obtained as the quotient of the number of enrollments $\mathit{NE}_{i,t}$ in state $i$ and the so-called eligible set $\mathit{EHG}_{i,t}$ of high school graduates for year $t$ coming to or staying in state $i$, which can generally differ substantially from the own-state high school graduates $\mathit{HG}_{i,t}$ in $i$ of this specific year. We set
\begin{align} \label{eq:y}
y_{i,t}= \dfrac{\mathit{NE}_{i,t}}{\mathit{EHG}_{i,t}} \ ,
\end{align}

\begin{figure}
	\centering
	\includegraphics[width=0.7\textwidth]{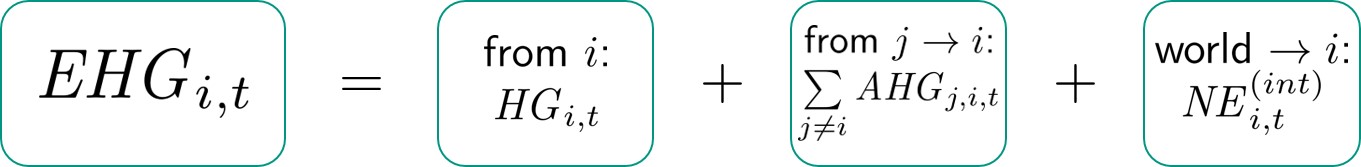}
	\caption{Illustration of the composition of the eligible set $\mathit{EHG}_{i,t}$.}
	\label{fig:affected_grads}
\end{figure}
where we model $\mathit{EHG}_{i,t}$ to consist of three main different groups, namely own $i$-specific high school graduates $\mathit{HG}_{i,t}$, ``affected'' graduates $\mathit{AHG}_{j,i,t}$ from other German states and the number of new international enrollments in $i$, $\mathit{NE}^{(int)}_{i,t}$ (see Figure \ref{fig:affected_grads}):
\begin{align}\label{eq:ehg}
	 \mathit{EHG}_{i,t}=\mathit{HG}_{i,t}+\sum_{j\neq i} \mathit{AHG}_{j,i,t}+ \mathit{NE}^{(int)}_{i,t} \ .
\end{align}
While respective enrollment numbers $\mathit{NE}^{(i)}_{i,t}$ from $i$ in $i$, $\mathit{NE}^{(j)}_{i,t}$ from $j$ to $i$ and $\mathit{NE}^{(int)}_{i,t}$ of international students in $i$ are publicly available for any state $i$ in WT $t/t+1$, there is, however, no available direct data for the respective eligible quantities in\eqref{eq:ehg}. For the from $i$ to $i$ component, this can be well approximated by its upper bound of the number of all high school graduates in $i$ as in the German federal system, the ``home state'' of the high-school diploma is often part of the immediate choice set of university entrants. Since the share of international students remains stable at around 15\% over the years due to effects such as language barriers in German undergraduate programs, we assume that the low amount of tuition fees in the international context has no effect and we therefore only use the lower bound $\mathit{NE}^{(int)}_{i,t}$ in the eligible set. Though for the eligible part of potential movers $\mathit{AHG}_{j,i,t}$ from $j$ to $i$ within Germany, extreme approximations by its lower bound of the number of enrollments $\mathit{NE}^{(j)}_{i,t}$ or the upper bound of all graduates $\mathit{HG}_{j,t}$ in $j$ are too coarse. In particular in view of  tuition fee interventions, it is clear that $\mathit{AHG}_{j,i,t}$ is affected, but unclear how. We therefore model it explicitly as a convex combination between the potential extremes.
\begin{align}\label{eq:ahg}
\mathit{AHG}_{j,i,t}=\theta NE_{i,t}^{(j)} + (1-\theta)\mathit{HG}_{j,t} \ ,
\end{align}
with $\theta \in [0,1]$. Of course, choosing $\theta$ too low, i.e. giving $\mathit{HG}_{j,t}$ too much influence, will yield $y_{i,t}$ values that are unrealistically low. An absolute lower boundary would be a mean enrollment of $\bar{y}_{0.90}=0.25$, which is achieved at $\theta=0.9$. Looking at the aggregated number of all new enrollments (not just first-time students) in all of Germany from German high schools over 2003-2014 divided by all high school graduations in Germany at that time in our data, we have a mean enrollment rate of around $0.72$, which can serve as a very rough proxy for where to expect realistic values. If we only look at first-time enrollments, the rates have monotonically increased from 40\% in 2009 over the years.\footnote{Data source: federal ministry of education (BMBF) data webspace \url{http://www.datenportal.bmbf.de/portal/de/K253.html} Table 1.9.3} We therefore take $\theta=0.98$ as a reasonable lower $\theta$-boundary, which yields $\bar{y}_{0.98}\approx 0.4$. We then conduct our analysis transparently over a grid of $\theta$-values in between 0.98 and 1 which we denote as admissible $\theta$s and which yield mean enrollment rates $\bar{y}_{\theta}\geq 0.4$. Figure \ref{fig:y_mean} in Appendix \ref{Sec:Fig_Tables} shows the mean enrollment rates over $\theta$ indicating the sensitivity of $y$ with respect to $\theta$ in the considered range.

With additional information using the number of new enrollments $\mathit{NO}_{j,t}$ with graduation in state $j$ enrolling anywhere in Germany at $t$ combined with $NE_{i,t}^{(j)}$ and $\mathit{HG}_{j,t}$ we can augment the approximation of $\mathit{EHG}_{i,t}$. Moreover, in order to additionally control for effects from postponers $\mathit{HG}_{t-1}, \mathit{HG}_{t-2}$  in $\mathit{EHG}_{i,t}$, we employ extra non-public information\footnote{Provided by the Federal Statistics Office on request for a fee.} on the number of new enrollments $\mathit{NE}^{(j)}_{\tau,i,t}$ in state $i$ in WT $t/t+1$ with high school diploma obtained in year $\tau$. With this, we can obtain an alternative approximation $\mathit{AHG}_{j,i,t}^{*}$ of the number of high school graduates in $j$ potentially moving to $i$ at $t$
\begin{align}\label{eq:WHSG_new}
\mathit{AHG}_{j,i,t}^{*}=\max\Bigg\{\sum_{l=0}^{2} c_{i,j,t,t-l} \mathit{HG}_{j,t-l} \ , \ \mathit{NE}_{i,t}^{(j)} \Bigg \} \ ,
\end{align}
with share $c_{i,j,t,\tau}= \dfrac{\mathit{NE}^{(j)}_{\tau,i,t}}{\mathit{NO}_{j,t}} $ of enrollments from $j$ to $i$ within the cohort of $t-l$ relative to all enrollments from $j$ in year $t$, approximating the potentially moving share of the graduates $\mathit{HG}_{j,\tau}$ (See Table \ref{Tab:Var_expl} and Figure \ref{fig:var_response} in Appendix \ref{Sec:Fig_Tables} for a (graphical) overview of involved sets and their role) .\footnote{As it can happen that $\mathit{NO}_{j,t}>\mathit{HG}_{j,t-l}, \ l=0,1,2$, we ensure that $\mathit{AHG}_{j,i,t}^{*}$ is at least $\mathit{NE}_{i,t}^{(j)}$.} We focus on numbers up to a time lag of $l=2$ in $\tau=t-l$, which cover generally more than 75\% of enrollments (on the German level), and use this  graduation time specific information also for state $i$ to get a refined approximation of $\mathit{EHG}_{i,t}$ by
\begin{align}\label{eq:ehgstar}
	\mathit{EHG}^*_{i,t}=\mathit{HG}_{i,t}+ \sum_{l=1}^{2}c_{i,i,t,t-l} \mathit{HG}_{i,t-l} +\sum_{j\neq i} \mathit{AHG}^*_{j,i,t}+ \mathit{NE}^{(int)}_{i,t} \ .
\end{align}
Note that for a choice of $\theta^*=0.9927$, the empirical mean squared and mean absolute deviation of $\mathit{EHG}^*_{i,t}$ and $\mathit{EHG}_{i,t}$ over all $i$ and $t$ are minimized and both almost coincide. As a robustness check to our pure public data analysis, we also report results for a response $y^{\mathit{extra}}_{i,t}=\dfrac{\mathit{NE}_{i,t}}{\mathit{EHG}^*_{i,t}}$.\\

In the covariates, we model the treatment effect $d_{i,t}$ of a tuition fee as a dummy, with $d_{i,t}=1$ indicating an existing tuition fee in state $i$ in the winter term starting in year $t$ and $d_{i,t}=0$ otherwise.\footnote{In Germany, there were no fees for students studying for their first degree in public institutions from WT of 2014 and onwards. Before that, the maximum amount for first degree studies was limited to \euro1000 per year. Almost all universities made use of the maximum amount, thus suggesting a dummy variable design.} Because of German laws, each state could strategically decide on the introduction and timing of fees.

\begin{figure}
	\begin{minipage}{0.48\textwidth}
		\centering
		\includegraphics[width=0.95\textwidth]{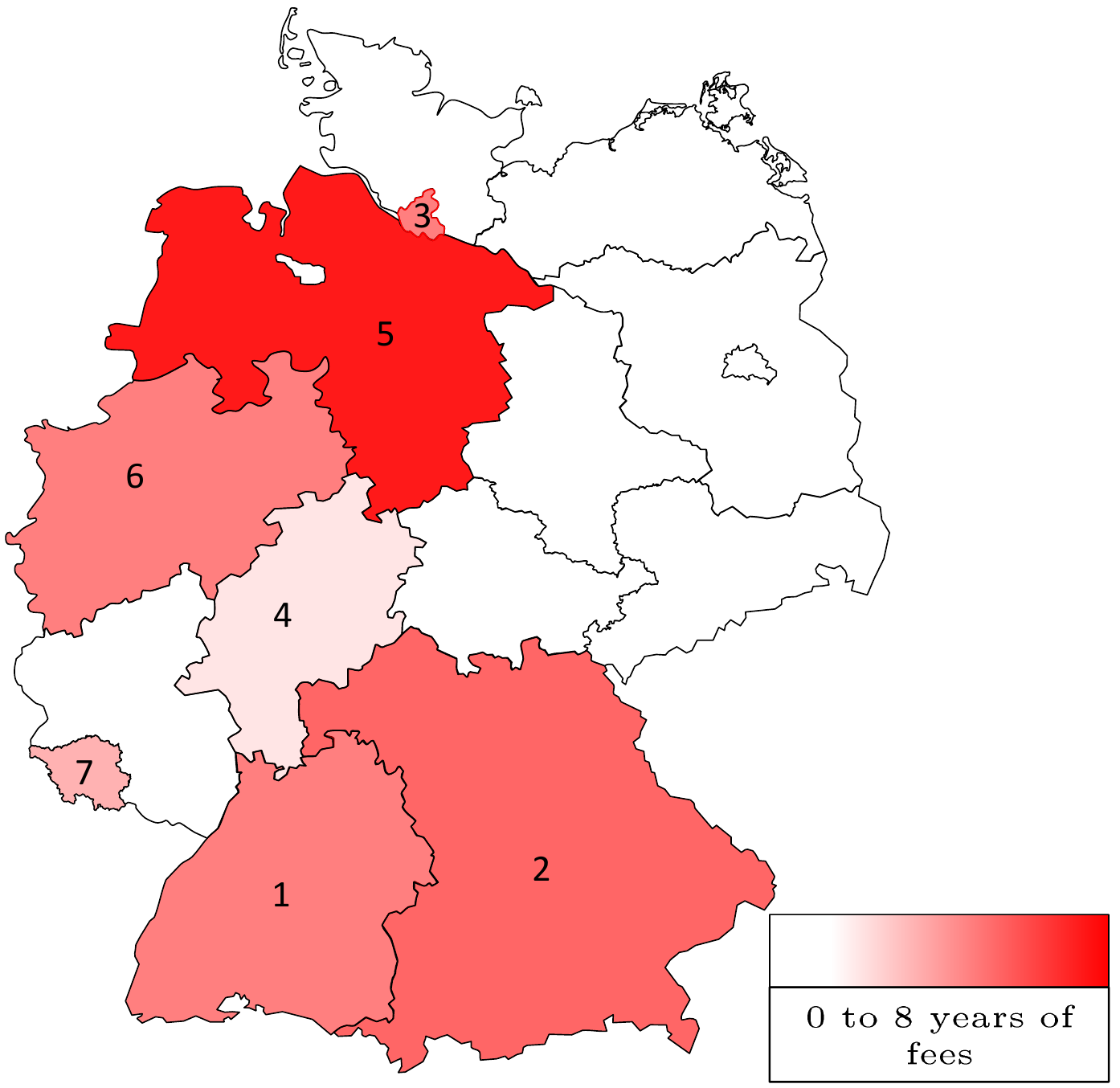}
	\end{minipage}\hfill
	\begin{minipage}{0.48\textwidth}
		\centering
		\includegraphics[width=0.95\textwidth]{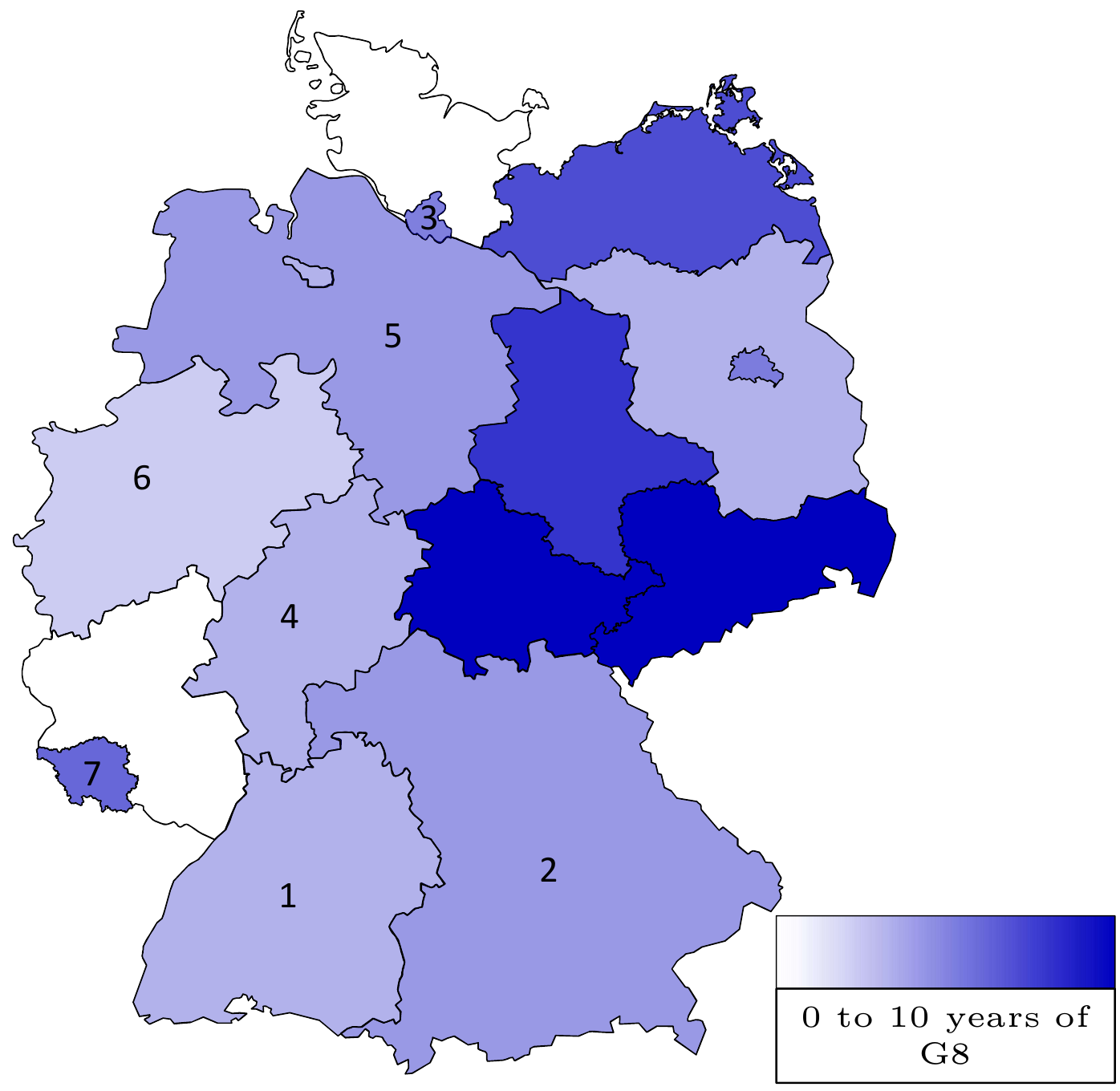}
	\end{minipage}\hfill
	
	\caption{Overview of the presence of tuition fees (\color{red}{left}\color{black}) and the G8-reform (\color{blue}{right}\color{black}) in the 16 German states until 2015. Darker colors represent longer presence of the respective variable.}
	\label{Bundeslaender}
\end{figure}

We generate spatial controls $z_{i,t}$ that capture migration behavior to each state from other state groups, which are formed depending on proximity and fees. This is necessary because of the heterogeneity of introduction and abolishment of tuition fees over states that can be seen in Figure \ref{fig:tuition_time}. Additionally, there are many cases where fee-states border non-fee states, which is highlighted in Figure \ref{Bundeslaender}. We therefore construct the spatial controls to measure the share of new enrollments in state $i$ that obtained their high school diploma in another state group. For each state $i$, we measure the proportion of new enrollments from a specific state group (e.g. neighboring fee states) relative to all enrollments in $i$. The groups consist of fee states that have a shared border with $i$, fee-states without a shared border with $i$, non-fee states, and enrollments from outside Germany (\textit{Migration.international}). For example, \textit{Migration.neighbor.fees} measures the proportion of new enrollments from all fee states with shared border to $i$ relative to all enrollments in $i$ that year. A detailed description can be found in Table \ref{Tab:Spatial_Controls} in Appendix \ref{Sec:Fig_Tables}.
Furthermore, to control for non-constant state specific effects, we employ 14 control variables $x_{i,t}$ using data from the socio-economic panel (SOEP)\footnote{We use the SOEP-long version 31. More information at \url{https://www.diw.de/en/diw_01.c.519381.en/1984_2014_v31.html}; for the usage, see \citet{SOEP}} and Destatis\footnote{More information at \url{https://www.destatis.de/EN}. Some variables were generated using data from Genesis-online database of Destatis accessible at \url{https://www-genesis.destatis.de}.}, the Federal Statistical Office in Germany. A detailed description can be found in Table \ref{Tab:Descriptive_Uni_Students} and Table \ref{Tab:Descriptive_socio} in Appendix \ref{Sec:Fig_Tables}. Together with the spatial variables, we have a set of $p=18$ potentially relevant covariates plus the binary variable of tuition fees.  Among others, we capture socio-economic variables comprised of urbanization level, income, rent, life satisfaction, unemployment rate and university and student related controls on staff and graduation statistics, the student-to-researcher ratio and data on the funding of universities. In particular, this set of variables contains all types of relevant controls from similar, previous studies (e.g. \citet{Bruckmeier2014,Mitze2015}). Moreover, we include two variables on the G8-reform that reduced the time of secondary education from nine to eight years. The implementation of this major educational policy change was also heterogeneous across states and is illustrated in \color{blue}{blue }\color{black} in Figure \ref{Bundeslaender}. This reform almost immediately substantially impacted the timing and the overall likelihood of much younger high school graduates to enroll to a university. We control for this effect with a dummy $\mathit{G8}_{i,t}$, where positive values indicate that the G8-reform was implemented in this state $i$, and additionally mark transition period years of double cohorts of G8 and G9 cohorts graduating by $\mathit{DC}_{i,t}=1$.
\begin{figure}
	\includegraphics[width=0.69\textwidth]{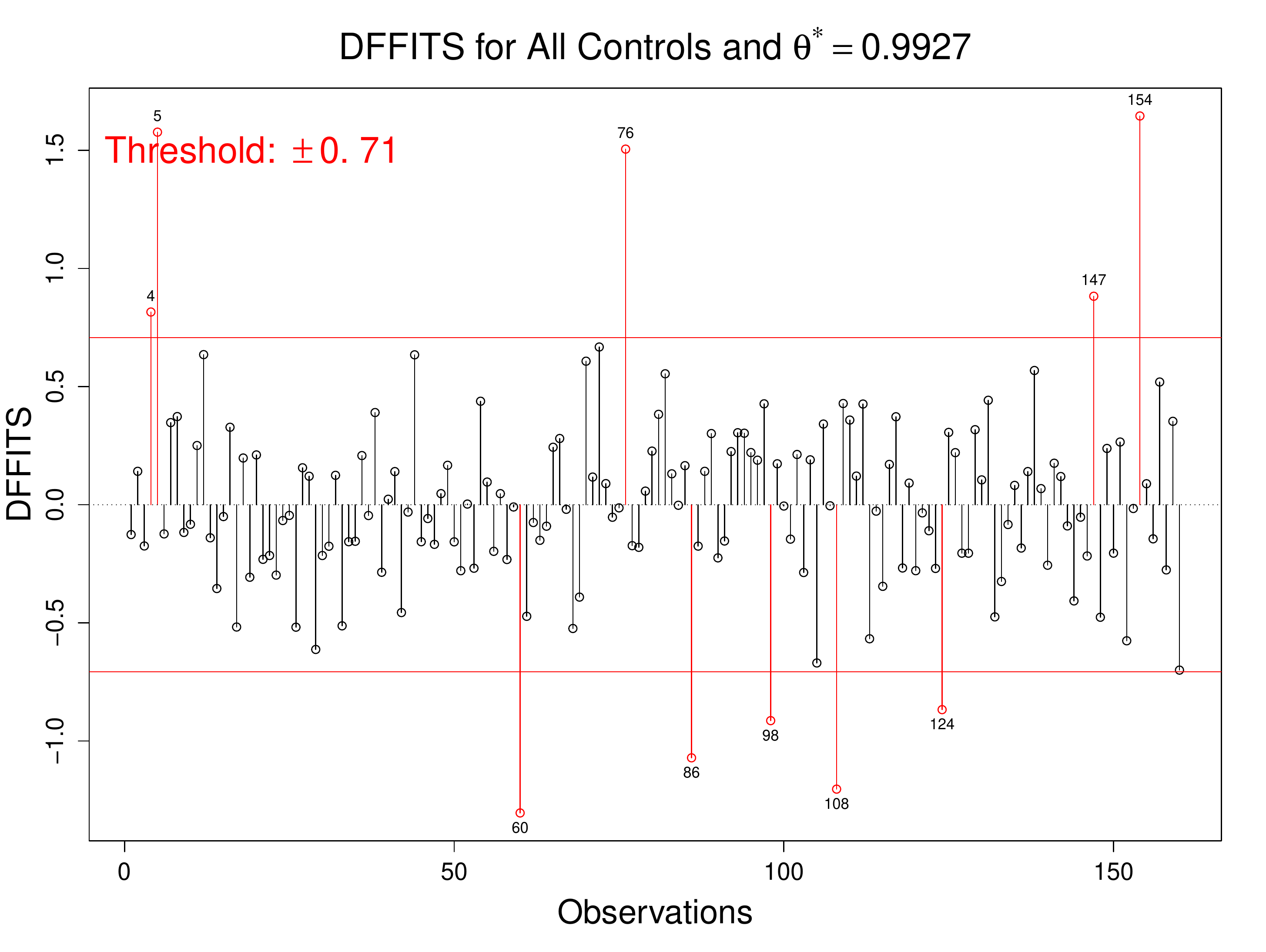}
	\includegraphics[width=0.3\textwidth]{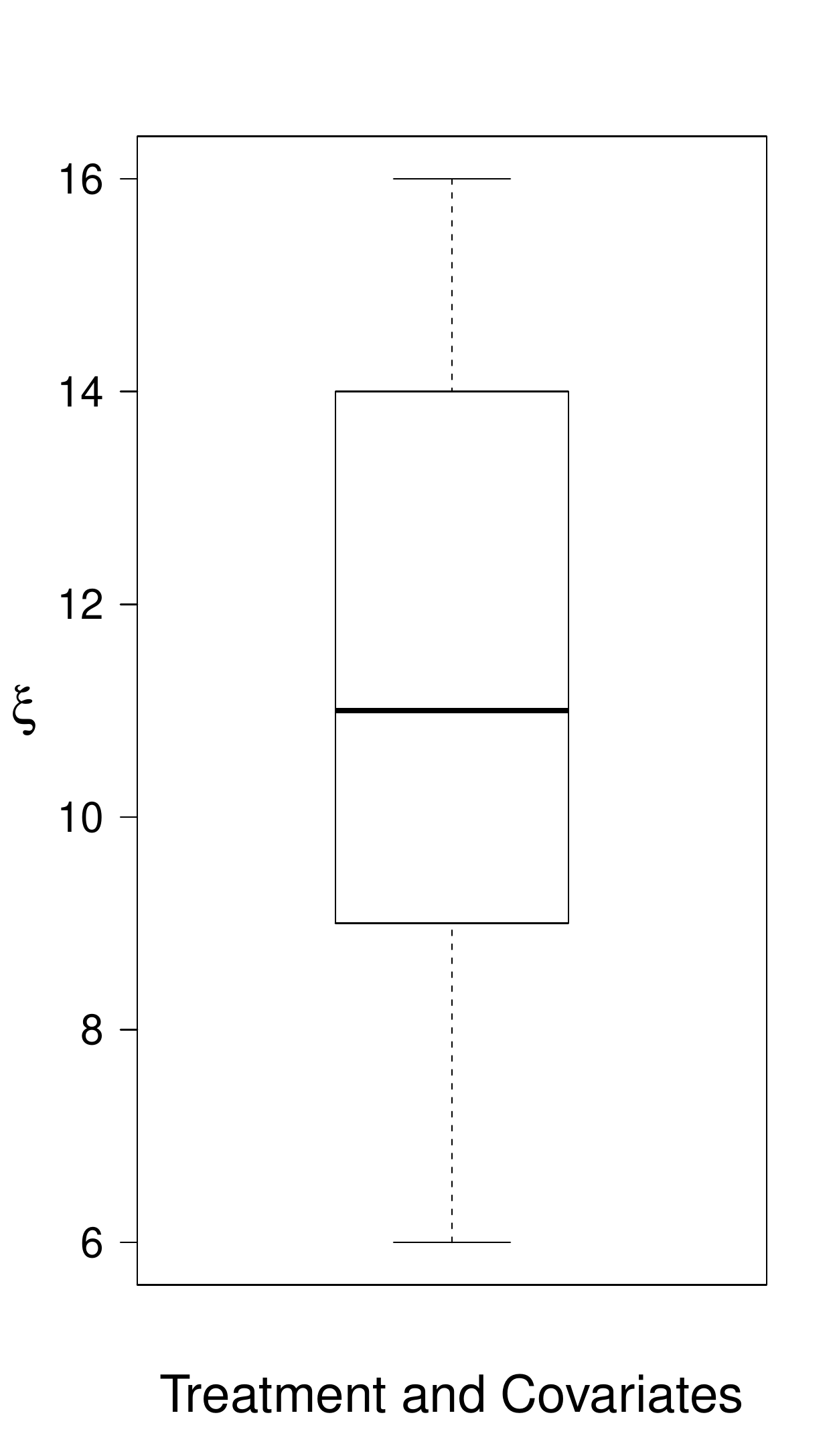}

\caption{Left: DFFITS for $\theta^{*}=0.9927$ and all controls: influential observations in \color{red}{red}\color{black}. Right: Boxplot of $\xi$ for the 18 covariates and the treatment with threshold: $\pm0.16$. Figures for each covariate available from authors upon request.}
\label{Fig:DFBETA}
\end{figure}

Inspecting the data, we find that single observations are highly influential. The left-hand side of Figure \ref{Fig:DFBETA} shows that the fitted enrollment rates heavily change when specific single observations are dropped from the regression estimation. More importantly, when looking at the leverage of covariates, we can inspect how coefficients change when these specific single observations are left out of the regression. If many influential observations affect one covariate, its selection by Lasso would depend strongly on these observations. The diagnostic tools used here are the DFFITS for changes in $y$ and the DFBETAS for changes in coefficients of covariates. Thresholds to decide whether or not observations are influential are calculated as $\pi_{\mathit{DFF}}=\frac{\sqrt{p}}{N}$ for DFFITS and $\pi_{\mathit{DFB}}=\frac{2}{\sqrt{N}}$ for DFBETAS, with $N=nT$ as the stacked number of observations. More specifically, with $g=1,\dots,p$, and $\mathit{DFB}_{g,k}$ as the DFBETAS measure of the $k$th observation of covariate $g$, let $\xi_g=\sum_{k=1}^{N}\mathds{1}_{\{\vert\mathit{DFB}_{g,k}\vert>\pi_{\mathit{DFB}}\}}$. $\xi_g$ therefore measures how many influential observations exist for each covariate $g$. The boxplot of $\xi$ in Figure \ref{Fig:DFBETA} shows that all covariates suffer from this phenomenon, indicating that the selection is unstable.
In addition to high expected correlations between regressors, it further encourages the use of stability selection instead of using all data points just once.

\section{Model and Methodology} \label{Section:model}

\subsection{Model} \label{sub:model}
The key goal of our study is to determine a finite sample precise estimate of the causal effect of tuition fees $\beta_{(0)}$ on enrollment rates $y$. For this, we employ standard identification assumptions to identify a constant causal impact in a linear panel set-up. Our contribution is the model determination with a stable but parsimonious data-driven selection of controls. For settings with limited data with potential measurement issues, this not only prevents cherry-picking of variables but also countervails biased causal effects for particular strong correlations of treatment and controls. Moreover, we illustrate how correct standard errors can be obtained quantifying the causal uncertainty when working with a complete population rather than a sample.

We use a two-equation linear panel model with fixed effects $\alpha_i$, where the covariates in both equations consist of socio-economic variables $x_{i,t}$ and spatial factors $z_{i,t}$. In the outcome equation, for each admissible $\theta$ in \eqref{eq:ahg}, the focus is on the linear causal effect of the tuition fee dummy $d_{i,t}$ on enrollments $y_{i,t}(\theta)$ given the large set of controls $(x_{i,t},z_{i,t})$.\footnote{For ease of exposition, we omit $\theta$ in the following in $y_{i,t}(\theta)$.} The auxiliary propensity score equation is also linear in $(x_{i,t},z_{i,t})$ and only serves as a correction device for data-driven model selection in the outcome equation due to correlation of  $d_{i,t}$ and $(x_{i,t},z_{i,t})$. Thus we work with the following model specification for $i=1,\dots,n=16$ states and $t=1,\dots,T=10$ years
\begin{eqnarray}
y_{i,t}&= &\beta_{(0)}d_{i,t} + \beta_{(1)}\transpose \binom{x_{i,t}}{z_{i,t}} + \alpha_i+ \epsilon^{(1)}_{i,t} \ ,\label{Model:unobserved}\\
d_{i,t}&= & \beta_{(2)}\transpose \binom{x_{i,t}}{z_{i,t}} + \epsilon^{(2)}_{i,t} \ ,\label{eq:auxiliary}
\end{eqnarray}
with  $y_{i,t}, \ \beta_{(0)}, \ d_{i,t}, \ \alpha_i, \, \ \epsilon^{(1)}_{i,t}, \ \epsilon^{(2)}_{i,t} \in \mathbb{R}$ and $\binom{x_{i,t}}{z_{i,t}} \in \mathbb{R}^p$ with $p=18$. Given the large set of controls including spatial factors for potential migration effects, the strict exogeneity conditions for both equations can be assumed as fulfilled, i.e. it holds that $ \mathbb{E}[\epsilon^{(1)}_{i,t} \mid d_{i,1},\dots,d_{i,T},x_{i,1},\dots,x_{i,T},z_{i,1},\dots,_{i,T},\alpha_i]=0 , \ \mathbb{E}[\epsilon^{(2)}_{i,t} \mid x_{i,1},\dots,x_{i,T},z_{i,1},\dots,z_{i,T}]=0 $. Note that $\alpha_i$ are fixed effects comprising e.g. unobserved regional aspects such as climate conditions, culture, or the topography of a state which might generally be correlated with at least some of the covariates $(x_{i,t},z_{i,t})$ such as e.g. rent or the urbanization level. Thus we work with the standard fixed effects transformation of \eqref{Model:unobserved} and \eqref{eq:auxiliary} removing $\alpha_i$ by demeaning:
\begin{eqnarray}
\ddot y_{i,t}&= &\beta_{(0)}\ddot d_{i,t} + \beta_{(1)}\transpose \binom{\ddot x_{i,t}}{\ddot z_{i,t}} + \ddot\epsilon^{(1)}_{i,t} \ ,\label{Model:unobserved2}\\
\ddot d_{i,t}&= & \beta_{(2)}\transpose \binom{\ddot x_{i,t}}{\ddot z_{i,t}} + \ddot\epsilon^{(2)}_{i,t} \ ,\label{eq:auxiliary2}
\end{eqnarray}
with $\ddot{y}_{i,t}=y_{i,t}-\overline{y}_{i}$ with $\overline{y}_{i}=\dfrac{1}{T}\sum_{t=1}^{T}y_{i,t}$ and similarly $\ddot{d}_{i,t}$, $\ddot{x}_{i,t}$, $\ddot{z}_{i,t}$, $\ddot{\epsilon}^{(1)}_{i,t}$, $\ddot{\epsilon}^{(2)}_{i,t}$.

Note that linearity in both equations is key for the identification of the causal effect $\beta_{(0)}$. Within this demeaned model, the linear form combined with the strict exogeneity ensures that the marginal effect $\beta_{(0)}$ of $d_{i,t}$ coincides with the average causal effect of $d_{i,t}$. This is easily seen by \eqref{Model:unobserved2} in a potential outcomes framework with $\ddot{y}_{i,t}(1)$ as the outcome when receiving the treatment $\ddot d_{i,t}$ and $\ddot{y}_{i,t}(0)$ when not receiving it. This implicitly incorporates the key identifying assumption of
\begin{align}\label{eq_id}
\mathbb{E}[\ddot{y}_{i,t}(0) \mid D_{i}, X_{i}, Z_{i}]&= {\beta}_{(1)}\transpose\binom{\ddot{x}_{i,t}}{\ddot{z}_{i,t}},
\end{align}
where $D$, $X$ and $Z$ are the stacked vectors of $\ddot{d}_{i,t}$, $\ddot{x}_{i,t}$ and $\ddot{z}_{i,t}$, i.e. $D_i=(\ddot{d}_{i,1},\dots, \ddot{d}_{i,T})\transpose$, $X_i=(\ddot{x}_{i,1},\dots, \ddot{x}_{i,T})\transpose$ and $Z_i=(\ddot{z}_{i,1},\dots, \ddot{z}_{i,T})\transpose$. Equation \eqref{eq_id} implies that $\mathbb{E}[\ddot{y}_{i,t}(0) \mid D_i, X_i, Z_i]=\mathbb{E}[\ddot{y}_{i,t}(0) \mid X_i, Z_i]$. This assumption is justified in our setting as decisions about the implementation of tuition fees in each state were taken at least one or two years ahead of the implementation date, and  were thus not influenced by actual enrollment numbers $y_{i,t}$. In our set-up, matching or propensity score estimates coincide with the marginal effects estimate for ${\beta}_{(0)}$ in \eqref{Model:unobserved2}. Here, the auxiliary equation \eqref{eq:auxiliary2} estimating the propensity score is only important to safeguard against underspecification from data-driven model choice in \eqref{Model:unobserved2} which would lead to biased estimates.

\subsection{Robust Model Selection and Post-Lasso Inference}
The proposed model selection and estimation procedure is two-step, where in step one, covariates are automatically selected separately in the outcome and the auxiliary equation. In step two, the union of the two sets of pre-selected covariates is then used to identify the causal effect of interest. Moreover, in our situation of $\frac{nT}{p}=8.89$, observations are so scarce relative to the dimensionality of the problem that plain OLS-type estimates are extremely imprecise. Thus for proper estimation of our main coefficient of interest $\beta_{(0)}$, we assume approximate sparsity, i.e., in fact only a few $s_{y}$ ($s_d$) of the other $p$ controls $x_{i,t}$  and $z_{i,t}$ are relevant for each state in the equation of $y$ ($d$).
We start from the reduced form of the main equation by plugging \eqref{eq:auxiliary2} into \eqref{Model:unobserved2}
\begin{equation}
\ddot y_{i,t}= \phi \transpose \binom{\ddot x_{i,t}}{\ddot z_{i,t}} + \ddot\eta_{i,t} \ ,\label{eq:reduced}\\
\end{equation}
with $\phi= \beta_{(1)}+ \beta_{(0)}\beta_{(2)}$ and $\ddot \eta_{i,t}= \ddot\epsilon^{(1)}_{i,t}+ \beta_{(0)}\ddot\epsilon^{(2)}_{i,t}$. We use the Lasso \citep{Tibshirani1996} as a data-driven tool to select the respective relevant covariates from an $\ell_1$ penalized minimization problem. We obtain the Lasso estimates $\hat{\beta}_{(1)}, \hat{\beta}_{(2)}$ as
\begin{align}
\label{lasso_outcome}
\hat{\beta}_{(1)}&=\underset{\phi}{\argmin}\ \frac{1}{2nT} \sum_{i=1}^{n}\sum_{t=1}^{T}\Bigg[ \ddot{y}_{i,t}- \phi\transpose \binom{\ddot{x}_{i,t}}{\ddot{z}_{i,t}} \Bigg]^2
+ \lambda_1 \sum_{j=1}^{p}\vert\phi^{(j)}\vert \ ,\\
\label{lasso_effect}
\hat{\beta}_{(2)}&=\underset{\beta_{(2)}}{\argmin}\ \frac{1}{2nT} \sum_{i=1}^{n}\sum_{t=1}^{T}\Bigg[ \ddot{d}_{i,t}-\beta_{(2)}\transpose \binom{\ddot{x}_{i,t}}{\ddot{z}_{i,t}} \Bigg]^2
+ \lambda_2 \sum_{j=1}^{p}\vert\beta_{(2)}^{(j)}\vert \ ,
\end{align}
with regularization parameters $\lambda_1,\lambda_2\geq0$ that are estimated by cross-validation and $\phi=(\phi^{(1)},\dots, \phi^{(p)})\transpose$ \footnote{In practice, there exist several techniques for solving this problem, while we use coordinate-descent algorithms \citep{Friedman2007,Friedman2010} provided in the \textit{glmnet} package in R.}. Note that we use the reduced form of the main equation~\eqref{eq:reduced} and therefore implicitly penalize the treatment also in~\eqref{lasso_outcome}. We use the lasso as a model selection device in both equations, where we denote the index set of selected covariates for \eqref{Model:unobserved2} by $S_y$ and for \eqref{eq:auxiliary2} by $S_d$. The causal effect can than be obtained from the post-selection equation using a union of both selected controls
\begin{align}\label{eq:postl}
{\ddot{y}}_{i,t}&= {\beta}_{(0)}\ddot{d}_{i,t} +{\tilde{\beta}}_{(1)}\transpose\binom{\ddot{x}_{i,t}^{S}}{\ddot{z}_{i,t}^{S}} + \ddot{\epsilon}^{(1)}_{i,t} \ ,
\end{align}
where $S=\hat{S}_y \cup \hat{S}_d \subseteq \{1,2,\dots,p\}$ , and $\ddot{x}^{S}_{i,t}$, $\ddot{z}^{S}_{i,t}$ only contain elements of $S$. Note that the  post-selection estimation in \eqref{eq:postl} is necessary in order to mitigate estimation biases from the penalized selection equations.

Instead of determining $\hat{S}_y$ and $\hat{S}_d$ as index set of elements in $(\ddot{x}_{i,t}\ddot{z}_{i,t})$ with non-zero $\hat{\beta}_{(1)}$ or $\hat{\beta}_{(2)}$ directly from \eqref{lasso_outcome} and \eqref{lasso_effect}(see \citet{Belloni2014a}), we suggest a subsampling-based stability selection. We demonstrate in the Section~\ref{Sec:Simulation} that this methodology also works for strongly correlated variables with measurement issues using the ideas and features of stability selection \citep{Meinshausen2010} in the Lasso selection steps \eqref{lasso_outcome} and \eqref{lasso_effect}. The procedure works as follows:
\begin{enumerate}
\item Generate $C$ subsamples $c$ of size $n^*$ of the $nT$ data points and obtain $C$ estimates $\hat{\beta}^{(j)}_{(1,c)}$ and $\hat{\beta}^{(j)}_{(2,c)}, \ c=1,\dots,C $ for each coefficient $j=1,\dots, p$ in \eqref{lasso_outcome} and \eqref{lasso_effect}.
\item Compute for each variable $j$ the relative inclusion frequencies $\hat{\Pi}_j^1=\frac{1}{C}{\sum_{c=1}^{1000}\mathds{1}_{\{\hat{\beta}^{(j)}_{(1,c)}\neq 0\}}}$ and $\hat{\Pi}_j^2=\frac{1}{C}{\sum_{c=1}^{1000}\mathds{1}_{\{\hat{\beta}^{(j)}_{(2,c)}\neq 0\}}}$.
\item Only include variable $j$ in the model and thus in $S$ if $\hat{\Pi}_j^1>\pi_{1}$ or $\hat{\Pi}_j^2>\pi_{2}$.
\end{enumerate}
Note that as in \citet{Belloni2014b,Belloni2014a}, $S$ consists of variables either influencing the treatment $d_{i,t}$ or the response $y_{i,t}$. Hence the selection choice from the auxiliary equation \eqref{lasso_effect} corrects wrong de-selection choices in the main enrollment equation \eqref{lasso_outcome} due to highly correlated control variables. In this sense it provides a robustification of the selection against underspecification and resulting biased estimates by double selection. In contrast to direct lasso in both selection equations, however, the proposed procedure reduces the risk of overspecification by the stability selection sub-sampling step.
Typically, the index set $S$ of the stability double selection is a subset of the standard double selected set and depends on the choice of sufficiently large $\pi_1$ and $\pi_2$ and the number of repetitions $C$. The stability post-double selection procedure yields a consistent $\beta_{(0)}$-estimator from \eqref{eq:postl}, see \citep{Belloni2014b, Meinshausen2010}. In contrast to standard lasso double selection, it also shows excellent finite sample performance in particular in settings with a very strong correlation of control variables in combination with single influential observations as in our data (see simulation study in Section~\ref{Sec:Simulation}).

For the empirical results and the simulation, we generally use $C=1000$ and $n^*=0.5nT$ in the algorithm above.\footnote{For the robustness checks using only the control year 2008 and 2014, we increase the subsample to $n^*=0.8nT$ to deal with the small data set.} For a data-driven threshold choice, we set minimum thresholds $\pi_{1,\theta}^{\mathit{min}},\pi_{2}^{\mathit{min}}>0.9$  as lower bounds to make ensuring that we screen out irrelevant variables. Since the response values change with $\theta$ in \eqref{lasso_outcome}, the corresponding minimum thresholds also depend on $\theta$. The selection of effective thresholds is then performed over a grid of threshold values starting from the minima increasing the threshold level to the first points where small changes in the thresholds do no longer change the model. The algorithm for the threshold choice can be found in Appendix \ref{Sec:Appendix_Threshold}. In the simulation, we also report estimates with $\pi_{1,\theta}^{\mathit{min}}=\pi_{2}^{\mathit{min}}=0.5$ and $0.7$ for comparison.

For inference, note that in our set-up we observe the full population, i.e. all states. Therefore uncertainty about the treatment effect does not result from sampling, but from uncertainty about the unobserved counterfactual. Thus instead of the usual HAC standard errors (\citet{MacKinnon1985}) we require standard errors that are specific to our set-up accounting for design uncertainty (see \citet{Abadie2020a}). We calculate such design-based standard errors $SE$ of the treatment effect $\beta_{(0)}$ in \eqref{eq:postl} as
\begin{equation}\label{eq:se}
  SE(\beta_{(0)})=\sqrt{(V\transpose V)^{-1}G(V\transpose V)^{-1}} \ ,
\end{equation}
where $V$ is the scalar residual from regressing $D$ jointly on $X$ and $Z$, and $G$ is the sample version of the variance $V_{\epsilon}$ of $\epsilon$ in \eqref{eq:postl}. $D$, $X$ and $Z$ are the stacked vectors of $\ddot{d}_{i,t}$, $\ddot{x}_{i,t}$ and $\ddot{z}_{i,t}$, i.e. $D=(\ddot{d}_{1,1},\dots, \ddot{d}_{i,t},\dots, \ddot{d}_{n,T})\transpose$, $X=(\ddot{x}_{1,1},\dots, \ddot{x}_{i,t},\dots,\ddot{x}_{n,T})$ and $Z=(\ddot{z}_{1,1},\dots, \ddot{z}_{i,t},\dots,\ddot{z}_{n,T})$. Details are in Appendix \ref{sec:appendix_se}. These standard errors are specific to the two equation post-selection estimation of $\beta_0$. They are consistent due to the linearity of both the outcome and the auxiliary propensity score equation (see \citet{Abadie2020a}, Assumption 8 and Theorem 1) while HAC standard errors are not. Moreover, for all statistical testing, we use the usual degrees of freedom ($df$) correction for fixed effects panel models.\footnote{The $df$ of the residuals reduce from $df=nT-\vert S \vert$ to $df=n(T-1)-\vert S\vert$, which is due to the demeaning process. For each observation $i$, one degree of freedom is lost because of the error term $\epsilon_{i,t}$. The latter is now comparable to a parameter that needs to be estimated (see \citet{Wooldridge2002}).}

\section{Simulation} \label{Sec:Simulation}
We conduct a Monte-Carlo Simulation to show the importance of stability selection when it is hard to disentangle effects of different covariates. This can be further adapted to our data by including influential observations and by inducing strong correlation among covariates. Using $i=1,\dots,n$, $t=1,\dots,T$, and $g=1,\dots,p$ with $T=10$, $n=16$, $N=nT$, and $p=30$, we simulate a linear panel model of the following form:
\begin{align*}
\tilde{y}_{i,t}&= \eta_0 d_{i,t} + \eta_1 \tilde{x}_{i,t} + \alpha_i+ \sigma_1(d_{i,t},x_{i,t})\epsilon^{(1)}_{i,t} \ ,\\
d_{i,t}&=  \eta_2 \tilde{x}_{i,t} +  \sigma_2(x_{i,t})\epsilon^{(2)}_{i,t} \ ,
\end{align*}
with coefficients depending on $g$: $\eta_0=0.5$, $\eta_1^{(g)}=\frac{5}{g}\mathds{1}_{\{g \leq 10\}}$, and $\eta_2^{(g)}=\frac{5}{g-6}\mathds{1}_{\{7 \leq g \leq 10\}}$ for $g\neq 6$, zero otherwise. The coefficients of covariates are up to 10 times higher than the coefficient of the treatment, since such large differences are also likely to arrive in our empirical application, where the expected treatment effect is relatively small. We generate the fixed effects as $\alpha_i \sim \mathcal{N}(0,\sqrt{\frac{4}{T}}) \ $ and $x_{i,t} \sim \mathcal{N}(0,\Sigma)$\footnote{$x_{i,t}=(x_{i,t}^{(1)},\dots,x_{i,t}^{(g)},\dots,x_{i,t}^{(p)})\transpose$: for $g,k=1,\dots,p$, $x_{i,t}^{(g)}$ represents a covariate that is standard normal with a correlation of $\rho=0.5^{k}$ to $x_{i,t}^{(g+k)}$ and $x_{i,t}^{(g-k)}$, $1 \leq g-k \leq g+k \leq p$.}, with $\Sigma_{v,w}=0.5^{\vert w-v\vert}$, $v$ representing the rows and $w$ the columns of $\Sigma$, $v\neq w$. For $v=w=1,\dots, 10$, $\Sigma_{v,w}=2$, and for $v=w=11,\dots, 30$, $\Sigma_{v,w}=6$. The errors are independently distributed as $\epsilon^{(1)}_{i,t} \sim \mathcal{N}(0,1)$ and $\epsilon^{(2)}_{i,t} \sim \mathcal{N}(0,1)$ with a heteroscedastic structure given by
\begin{align*}
\sigma_1(d_{i,t},x_{i,t})=\sqrt{\dfrac{(1+\eta_0 d_{i,t} +\eta_1 x_{i,t} + \alpha_i)^2}{\mathbb{E}_{N}[(1+\eta_0 d_{i,t} +\eta_1 x_{i,t} + \alpha_i)^2]}},\  \sigma_2(x_{i,t})=\sqrt{\dfrac{(1+\eta_2 x_{i,t})^2}{\mathbb{E}_{N}[(1+\eta_2 x_{i,t})^2]}} \ .
\end{align*}
Given this structure, we distort the last $10\%$ of observations by a vector $\gamma=(\gamma_1,\dots,\gamma_p)\transpose$, and we generate each $\gamma_{g}\sim U[\frac{2}{3}\mathit{inf},\mathit{inf}]$, where $\mathit{inf}\in \{0,1,5\}$ and $g \in \mathcal{D}$ depending on the scenario. In each scenario (i.e. different \textit{inf}-values), we distort covariates either from the active set ($\mathcal{D}= \{j: \ \vert\eta_1^{(j)}\vert+ \vert\eta_2^{(j)}\vert \neq 0 \}$), the inactive set ($\mathcal{D}= \{j: \ \vert\eta_1^{(j)}\vert+ \vert\eta_2^{(j)}\vert = 0 \}$) or the response $y$. For distortion of covariates, we modify them to $\tilde{x}_{i,t}=x_{i,t}+\gamma, \ t=10$. This means that $\gamma_g=0$ for either $g >10$ (inactive set) or $g\leq 10$ (active set). When $y$ is distorted, we have $\tilde{y}_{i,t}=y_{i,t}+ \zeta, \ t=10$ and $\zeta \sim U[-\mathit{inf},\mathit{inf}]$. We report mean values over 1000 replications for the absolute bias of estimators $\hat{\eta}_0$ from $\eta_0$, the root mean squared error for $\eta_0$ with $RMSE_{\eta_0}=\sqrt{\mathit{Bias}_{\eta_0,\hat{\eta}_0}^2 + \mathit{Var}_{\hat{\eta}_0}}$, the number of selected covariates, the true positive rate TPR=$\dfrac{\sum_{g=1}^{p} \mathds{1}_{\{\eta_1^{(g)}\neq 0\} }\mathds{1}_{\{\hat{\eta}_1^{(g)}\neq 0\}}}{\sum_{g=1}^{p} \mathds{1}_{\{\eta_1^{(g)}\neq 0\} }}$, and the false positive rate FPR=$\dfrac{\sum_{g=1}^{p} \mathds{1}_{\{\eta_1^{(g)}= 0\} }\mathds{1}_{\{\hat{\eta}_1^{(g)}\neq 0\}}}{\sum_{g=1}^{p} \mathds{1}_{\{\eta_1^{(g)}= 0\} }}$. We also report the rejection rate, which is based on conventional t-tests on the estimated $\hat{\eta}_0$ against the true $\eta_0$. For the t-tests and the $RMSE_{\eta_0}$, we use the suggested standard errors of \citet{Abadie2020a}. We additionally report results using the classical heteroscedasticity consistent standard errors \citep{MacKinnon1985} in Appendix \ref{Sec:Fig_Tables}. Results only change considering the rejection rates and the $RMSE_{\eta_0}$, where the classical HC3-standard errors are more conservative, resulting in smaller rejection rates and larger $RMSE_{\eta_0}$-values than their design-based counterparts. We report results from post-Lasso and post-double selection as the described in Section \ref{Section:model}, using no subsampling at all and using the subsampling similar to stability selection with $\pi_{\mathit{min}}\in \{0.5, \ 0.7\}$. Additionally, we report the two extreme cases using all covariates without selection (Fixed Effects all) and using only the true influencing variables (Oracle).

\afterpage{
	\clearpage
	\begin{landscape}
		\begin{table}[!htbp] \centering
			\captionof{table}{Simulation Results for Different Forms and Strengths of Influential Observations with Design-based Standard Errors}
			\label{Tab:Sim_results_abadie}
			\resizebox{1.35\textheight}{!}{%
				\begin{tabular}{
						l@{\hskip 20pt}ccc@{\hskip 30pt} ccc@{\hskip 30pt} ccc@{\hskip 30pt} ccc@{\hskip 30pt} ccc@{\hskip 30pt} ccc@{\hskip 30pt}}
					\\[-1.8ex]\hline
					\hline \\[-1.8ex] & \multicolumn{3}{c@{\hskip 30pt}}{Absolute $\mathit{Bias}_{\eta_0}$} & \multicolumn{3}{c@{\hskip 30pt}}{$\mathit{RMSE}_{\eta_0}$} & \multicolumn{3}{c@{\hskip 30pt}}{$\#$ Covariates} & \multicolumn{3}{c@{\hskip 30pt}}{TPR} & \multicolumn{3}{c@{\hskip 30pt}}{FPR} & \multicolumn{3}{c@{\hskip 30pt}}{Rejection Rate}   \\
					\cmidrule(r{30pt}){2-4} \cmidrule(r{30pt}){5-7} \cmidrule(r{30pt}){8-10} \cmidrule(r{30pt}){11-13} \cmidrule(r{30pt}){14-16} \cmidrule(r{30pt}){17-19} \\[-1.8ex]
					\textit{Size of Distortion:} & 0 & 1 & 5 & 0 & 1 & 5 & 0 & 1 & 5 & 0 & 1 & 5 & 0 & 1 & 5 & 0 & 1 & 5\\ \midrule \\[-1.8ex]
					\multicolumn{3}{l}{\textit{Distortion in the Active Set}}  \\
					PL Stab: 0.5 & 0.155 & 0.158 & 0.171 & 0.025 & 0.026 & 0.030 & 9.284 & 9.333 & 8.131 & 0.839 & 0.838 & 0.784 & 0.045 & 0.048 & 0.015 & 0.997 & 0.997 & 0.994 \\
					DB Stab: 0.5 & 0.087 & 0.087 & 0.081 & 0.020 & 0.020 & 0.018 & 10.680 & 10.697 & 10.144 & 1.000 & 1.000 & 1.000 & 0.034 & 0.035 & 0.007 & 0.121 & 0.128 & 0.103 \\
					PL Stab: 0.7 & 0.161 & 0.163 & 0.181 & 0.027 & 0.028 & 0.033 & 8.379 & 8.341 & 7.384 & 0.804 & 0.801 & 0.730 & 0.017 & 0.017 & 0.004 & 1.000 & 1.000 & 1.000 \\
					DB Stab: 0.7 & 0.087 & 0.086 & 0.081 & 0.020 & 0.020 & 0.018 & 10.187 & 10.199 & 10.027 & 1.000 & 1.000 & 1.000 & 0.009 & 0.010 & 0.001 & 0.126 & 0.114 & 0.106 \\
					Post Lasso & 0.144 & 0.143 & 0.121 & 0.023 & 0.023 & 0.021 & 19.377 & 19.193 & 16.813 & 0.917 & 0.918 & 0.935 & 0.510 & 0.501 & 0.373 & 0.886 & 0.866 & 0.610 \\
					Double Selection & 0.091 & 0.091 & 0.086 & 0.020 & 0.020 & 0.019 & 21.272 & 20.694 & 17.371 & 1.000 & 1.000 & 1.000 & 0.564 & 0.535 & 0.369 & 0.169 & 0.167 & 0.135 \\
					Fixed Effects All & 0.092 & 0.094 & 0.089 & 0.020 & 0.020 & 0.019 & 30.000 & 30.000 & 30.000 & 1.000 & 1.000 & 1.000 & 1.000 & 1.000 & 1.000 & 0.188 & 0.197 & 0.182 \\
					Oracle & 0.086 & 0.086 & 0.081 & 0.020 & 0.020 & 0.018 & 10.000 & 10.000 & 10.000 & 1.000 & 1.000 & 1.000 & 0.000 & 0.000 & 0.000 & 0.121 & 0.109 & 0.105 \\
					\\[-1.8ex]
					\multicolumn{3}{l}{\textit{Distortion in the Inactive Set}}   \\
					PL Stab: 0.5 & 0.155 & 0.157 & 0.157 & 0.025 & 0.026 & 0.026 & 9.284 & 9.403 & 9.249 & 0.839 & 0.840 & 0.842 & 0.045 & 0.050 & 0.042 & 0.997 & 0.994 & 0.996 \\
					DB Stab: 0.5 & 0.087 & 0.087 & 0.087 & 0.020 & 0.020 & 0.020 & 10.680 & 10.774 & 10.589 & 1.000 & 1.000 & 1.000 & 0.034 & 0.039 & 0.029 & 0.121 & 0.124 & 0.125 \\
					PL Stab: 0.7 & 0.161 & 0.163 & 0.162 & 0.027 & 0.028 & 0.027 & 8.379 & 8.369 & 8.300 & 0.804 & 0.801 & 0.802 & 0.017 & 0.018 & 0.014 & 1.000 & 1.000 & 1.000 \\
					DB.Stab: 0.7 & 0.087 & 0.086 & 0.086 & 0.020 & 0.020 & 0.020 & 10.187 & 10.234 & 10.168 & 1.000 & 1.000 & 1.000 & 0.009 & 0.012 & 0.008 & 0.126 & 0.122 & 0.119 \\
					Post Lasso   & 0.144 & 0.143 & 0.145 & 0.023 & 0.023 & 0.024 & 19.377 & 19.236 & 18.530 & 0.917 & 0.917 & 0.917 & 0.510 & 0.503 & 0.468 & 0.886 & 0.868 & 0.880 \\
					Double Selection & 0.091 & 0.092 & 0.092 & 0.020 & 0.020 & 0.020 & 21.272 & 21.061 & 20.537 & 1.000 & 1.000 & 1.000 & 0.564 & 0.553 & 0.527 & 0.169 & 0.171 & 0.178 \\
					Fixed Effects All & 0.092 & 0.094 & 0.094 & 0.020 & 0.021 & 0.021 & 30.000 & 30.000 & 30.000 & 1.000 & 1.000 & 1.000 & 1.000 & 1.000 & 1.000 & 0.188 & 0.195 & 0.200 \\
					Oracle & 0.086 & 0.087 & 0.087 & 0.020 & 0.020 & 0.020 & 10.000 & 10.000 & 10.000 & 1.000 & 1.000 & 1.000 & 0.000 & 0.000 & 0.000 & 0.121 & 0.114 & 0.114 \\
					\\[-1.8ex]
					\multicolumn{3}{l}{\textit{Distortion in the Response}}   \\
					PL Stab: 0.5 & 0.155 & 0.158 & 0.169 & 0.025 & 0.026 & 0.030 & 9.284 & 9.311 & 8.453 & 0.839 & 0.835 & 0.768 & 0.045 & 0.048 & 0.039 & 0.997 & 0.999 & 0.999 \\
					DB Stab: 0.5 & 0.087 & 0.088 & 0.108 & 0.020 & 0.020 & 0.031 & 10.680 & 10.799 & 10.786 & 1.000 & 1.000 & 1.000 & 0.034 & 0.040 & 0.039 & 0.121 & 0.125 & 0.111 \\
					PL Stab: 0.7 & 0.161 & 0.164 & 0.175 & 0.027 & 0.028 & 0.032 & 8.379 & 8.384 & 7.563 & 0.804 & 0.799 & 0.728 & 0.017 & 0.020 & 0.014 & 1.000 & 1.000 & 1.000 \\
					DB.Stab: 0.7 & 0.087 & 0.088 & 0.106 & 0.020 & 0.020 & 0.030 & 10.187 & 10.240 & 10.211 & 1.000 & 1.000 & 1.000 & 0.009 & 0.012 & 0.011 & 0.126 & 0.120 & 0.101 \\
					Post Lasso & 0.144 & 0.144 & 0.149 & 0.023 & 0.024 & 0.027 & 19.377 & 19.334 & 19.020 & 0.917 & 0.916 & 0.908 & 0.510 & 0.509 & 0.497 & 0.886 & 0.877 & 0.842 \\
					Double Selection & 0.091 & 0.092 & 0.111 & 0.020 & 0.021 & 0.031 & 21.272 & 21.164 & 21.497 & 1.000 & 1.000 & 1.000 & 0.564 & 0.558 & 0.575 & 0.169 & 0.170 & 0.143 \\
					Fixed Effects All & 0.092 & 0.095 & 0.114 & 0.020 & 0.021 & 0.031 & 30.000 & 30.000 & 30.000 & 1.000 & 1.000 & 1.000 & 1.000 & 1.000 & 1.000 & 0.188 & 0.192 & 0.156 \\
					Oracle & 0.086 & 0.087 & 0.106 & 0.020 & 0.020 & 0.030 & 10.000 & 10.000 & 10.000 & 1.000 & 1.000 & 1.000 & 0.000 & 0.000 & 0.000 & 0.121 & 0.115 & 0.102 \\
					\hline
			\end{tabular} }
			\captionof*{table}{\scriptsize{Note: All values are based on Monte Carlo simulations with 1000 runs and 1000 repeated subsample steps ($C=1000$). Rejection rates are based on t-tests with design-based standard errrors (see \citet{Abadie2020a}). The remaining measures are means over the 1000 replication runs. PL Stab and DB Stab stand for post-Lasso and double selection with stability selection and the corresponding minum thresholds $\pi_{\mathit{min}}$. Oracle is similar to Fixed Effects All but using only true influencing covariates. \textit{inf} indictates the strength of influential observations and is reported for each measure, while the form of influence (active/inactive set and response) is depicted in the rows.}}
		\end{table}
\end{landscape} }

Table \ref{Tab:Sim_results} summarizes our simulation results. First of all, as expected, the proposed double selection procedure combined with stability selection performs best overall and is almost identical to the oracle procedure that knows the true active set. Using of $\pi_{\mathit{min}}=0.7$ or $\pi_{\mathit{min}}=0.5$ does not affect results much in most cases. When distorting the inactive set, using a higher minimum threshold reduces the FPR even more than in other cases, as the noise variables have more influence. When regarding post-Lasso, however, $\pi_{\mathit{min}}=0.5$ seems to perform better in general, which can be explained by the post-Lasso not detecting all relevant covariates in the simulated data, where a lower threshold leads to the inclusion of more relevant variables compared to noise variables and improves the method here. For the double selection, only more noise variables are added since all relevant variables are already (almost) always detected. When distorting the response, bias and \textit{RMSE} values go up in general for all procedures, but their relative performance compared to the oracle does not get worse. Comparing stability procedures to their non-stable counterparts, we see that the latter include up to twice as many covariates without much improvement on the TPR, but high increases in the FPR. This confirms the hypothesis that without stability selection, many irrelevant covariates are included in the model, which increases the bias and \textit{RMSE}. The rejection rate is especially high for all post-Lasso procedures, which is not surprising given their high bias and relatively low standard errors that are a result of including fewer variables in the model. Small standard errors also affect the \textit{RMSE} values, and in scenarios with high distortions in the response y, the post-Lasso has a similar \textit{RMSE} compared to its double selection counterpart (regarding the stability procedures).

Taking a closer look at the different forms of distortion, we do not observe much change for high $\mathit{inf}$-values when we distort variables from the inactive set. As expected, when influential observations are only present in the noise variables, they do not affect the selection procedures much. When distorting the active set only, however, procedures with the post-Lasso select fewer (relevant) variables due to the added noise, which leads to a higher bias (for the stability cases), and increases \textit{RMSE} values. The double selection procedures seem to be very robust against such distortions, with all measures remaining relatively unchanged. This is not surprising, since the double selection procedure helps to reduce such a bias by taking the second equation into account. Finally, distorting the response is interesting, since both relevant and irrelevant covariates are affected at the same time. Even with extremely high distortions, the double selection procedures keep a lower bias compared to the other methods and double selection with stability selection has very low FPRs, while selecting almost all variables from the active set. All in all, the simulation shows that only when we use stability selection, we can select the right variables without including too many noise variables. In our simulated model, where it is hard to distinguish between covariates and the treatment effect is relatively small compared to the effects of other covariates, the non-stable methods perform worse over all distortion scenarios\footnote{Results are similar using a lower correlation among covariates. Additional simulations are available upon request.}. Furthermore, we see that when some covariates explain the treatment well, but only have a moderate effect on the response (which is the case in the application), double selection outperforms the post-Lasso in terms of bias and rejection rate.

\section{Empirical Results} \label{Sec:Empirical}
\subsection{Main Findings}
In this section, we present the results of our empirical study. Generally, with only publicly available data and the proposed post stability double selection methodology, we find that tuition fees in Germany significantly reduced the enrollment rate by 3.8pp to up to 4.5pp on average over all possible cases of response variables. For all admissible values of $\theta$, the procedure consistently identifies the same one university specific and one educational policy change control variable in $x$ and the four spatial variables $z$ as important drivers highlighting the importance of fee induced migration effects. Moreover, we find that during the considered period, other socio-economic factors only played a minor role. Given the transparency in $\theta$ and the data-driven stability double selection, we judge these findings are very robust.

\begin{table}[h]
	\centering
	\caption{Estimates of the Causal Effect of Tuition Fees $\beta_{(0)}$ for Different $\theta$-Values}
	
	\resizebox{0.9\textwidth}{!}{
		
		\begin{tabular}{@{\extracolsep{5pt}} cccccc}
			                                                       &                                        &                                        &                                        &  &  \\[-1.8ex] \hline\hline
			                                                       &                                        &                                        &                                        &  &  \\
			[-1.8ex]                                               &                                     \multicolumn{3}{c}{Double Selection + Stability}                                     &  &    \multicolumn{1}{c}{All Controls}    \\ \cline{2-5}\cline{6-6}
			Effects on $y_{i,t}$                                   &             $\theta=0.98$              &           $\theta^*=0.9927$            &               $\theta=1$               &  &           $\theta^*=0.9927$            \\ \hline
			                                                       &                                        &                                        &                                        &  &  \\[-1.8ex]
			\textbf{Tuition Fees} & $\mathbf{-4.310}$ & $\mathbf{-3.996}$ & $\mathbf{-3.808}$ &  & $\mathbf{-1.267}$ \\
			   $\underset{(\textit{HC3})}{\textit{(Design-based)}}$ & $\mathit{ \underset{(1.243)}{(1.177)}}$ & $\mathit{ \underset{(1.372)}{(1.278)}}$ & $\mathit{ \underset{(1.593)}{(1.486)}}$ &  & $\mathit{ \underset{(1.229)}{(0.989)}}$ \\
			   &                                        &                                        &                                        &  &  \\[-1.8ex]
			\color{kit-blue}{Student.to.researcher.ratio}          &                $-2.763$                &                $-2.931$                &                $-3.286$                &  &                $0.887$                 \\
			\color{kit-blue}{Double.Cohort}                        &                $-1.732$                &                $-2.766$                &                /                       &  &                $-6.294$                \\
			Migration.neighbor.fees                                &                   /                    &                $46.443$                &                $86.812$                &  &                $31.254$                \\
			Migration.rest.fees                                    &                $59.847$                &                $83.003$                &               $134.509$                &  &                $71.765$                \\
			\color{kit-blue}{Migration.international}                                &                $-3.958$                &                $15.621$                &                $35.069$                &  &               $-21.623$                \\
			Migration.no.fees                                      &                $22.956$                &                $46.713$                &                $77.884$                &  &                $30.678$                \\
			\vdots                                                 &                   /                    &                   /                    &                   /                    &  &                 \vdots                 \\ \hline\hline
		\end{tabular} }
	\caption*{\scriptsize{Note: Response values are scaled to a percentage level. Standard errors in parentheses are calculated based on treatment design and finite populations (\citet{Abadie2020a}) or heteroscedasticity consistent based on infinite populations(HC3, see \citet{MacKinnon1985}). Variables in \color{kit-blue}{blue} \color{black} appeared similarly in previous studies (not necessarily together).}}	
	\label{Tab:theta_098}
\end{table}

\begin{figure}[h]
	\centering
	\includegraphics[width=0.9\textwidth]{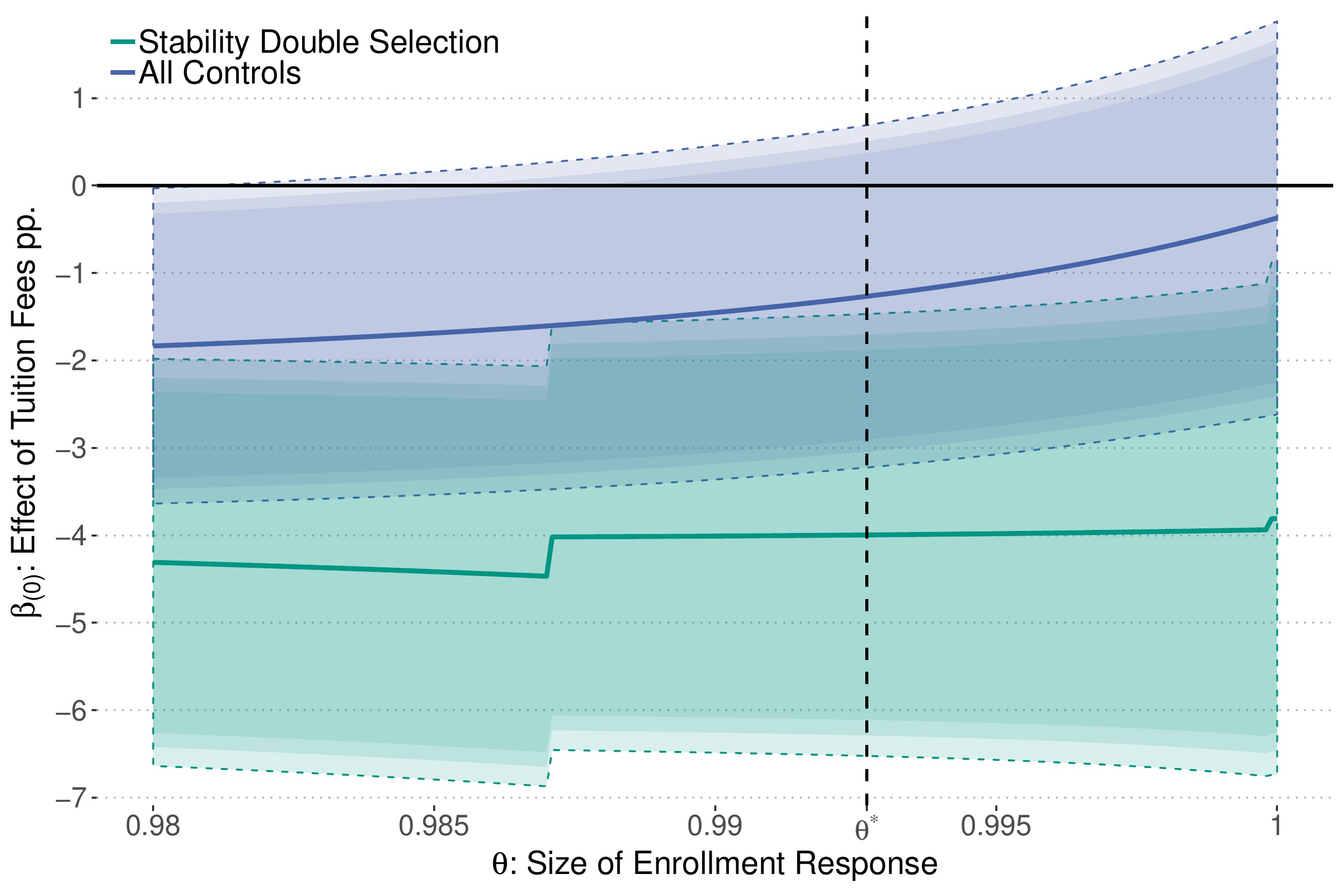}
	
	\caption{We report estimates for the causal effect $\beta_0$ in \eqref{Model:unobserved} for Stability Double Selection and using all controls in a linear fixed effects regression over the grid of admissible $\theta$ in $\mathit{AHG}_{j,i,t}$ from \eqref{eq:ahg}. We depict 95\%, 92.5\% and 90\% CIs in shaded colors, which are calculated on design-based standard errors (\citet{Abadie2020a}).}
	\label{Fig:grid}
\end{figure}
Table \ref{Tab:theta_098} summarizes the post-selection estimation results. Most importantly, we find a significant negative causal effect over the whole grid of $\theta$-values only when using post-double selection with repeated subsampling (Double Selection + Stability). The reference point $\theta^*=0.9927$ from additional non-public information in \eqref{eq:ehgstar} suggests in fact that values very close to the right boundary of $\theta=1$ are the most plausible, i.e. the number of effective enrollments of migrating students from $j$ to $i$ within Germany almost coincides with the number of potentially enrolling ones $\mathit{EHG}_{i,t}$ at $\theta^*$. For such large $\theta$-values in particular, using all controls in a plain panel OLS clearly underestimates the effect and thus leads to inflated p-values, which is illustrated in Figure \ref{Fig:grid}. Post-double selection Lasso without the stabilizing subsampling does not work as it leads to the same results as a pooled OLS with all controls. In those cases, the magnitude of the effect from tuition fees is roughly four times smaller than for the post stability double selection and the impact becomes insignificant. Across all admissible $\theta$, only about a third of the controls are selected with our proposed procedure, which indicates that many plausible  controlling factors from the literature are in fact not relevant and dominated in this period of heterogeneous changes in educational policies across states.

Looking more closely at Figure \ref{Fig:grid}, we see that over the entire grid of admissible $\theta$-values, only the double selection procedure with subsampling guarantees good performance, whereas with all controls the estimated effect for $\beta_0$ vanishes with $\theta$ approaching the upper bound 1. With an effect of tuition fees close to zero for the upper $\theta$-boundary, and only half the size of the one by the stable double selection at the lower $\theta$-boundary, the pooled OLS appears biased in detecting individual influences in this situation, where observations are scarce relative to the dimension of the model. This behavior is not surprising, as many irrelevant controlling factors that might be spuriously correlated with the response and the treatment are present without selection. This is more critical at the upper $\theta$-boundary, where the variability of the response is higher. Furthermore, using the post-Lasso, even with stability selection, gives less stable and often insignificant results. The insignificance can be traced back to the lack of additional controls that are only added in the second step of the double selection procedure, whereas the rather unstable results can furthermore be accounted for by the difference in the selection procedure in the first step that includes the treatment in the equation. All this emphasizes the importance of using a post stability double selection as proposed.

\begin{figure}[h]
	\centering
	\includegraphics[width=\textwidth]{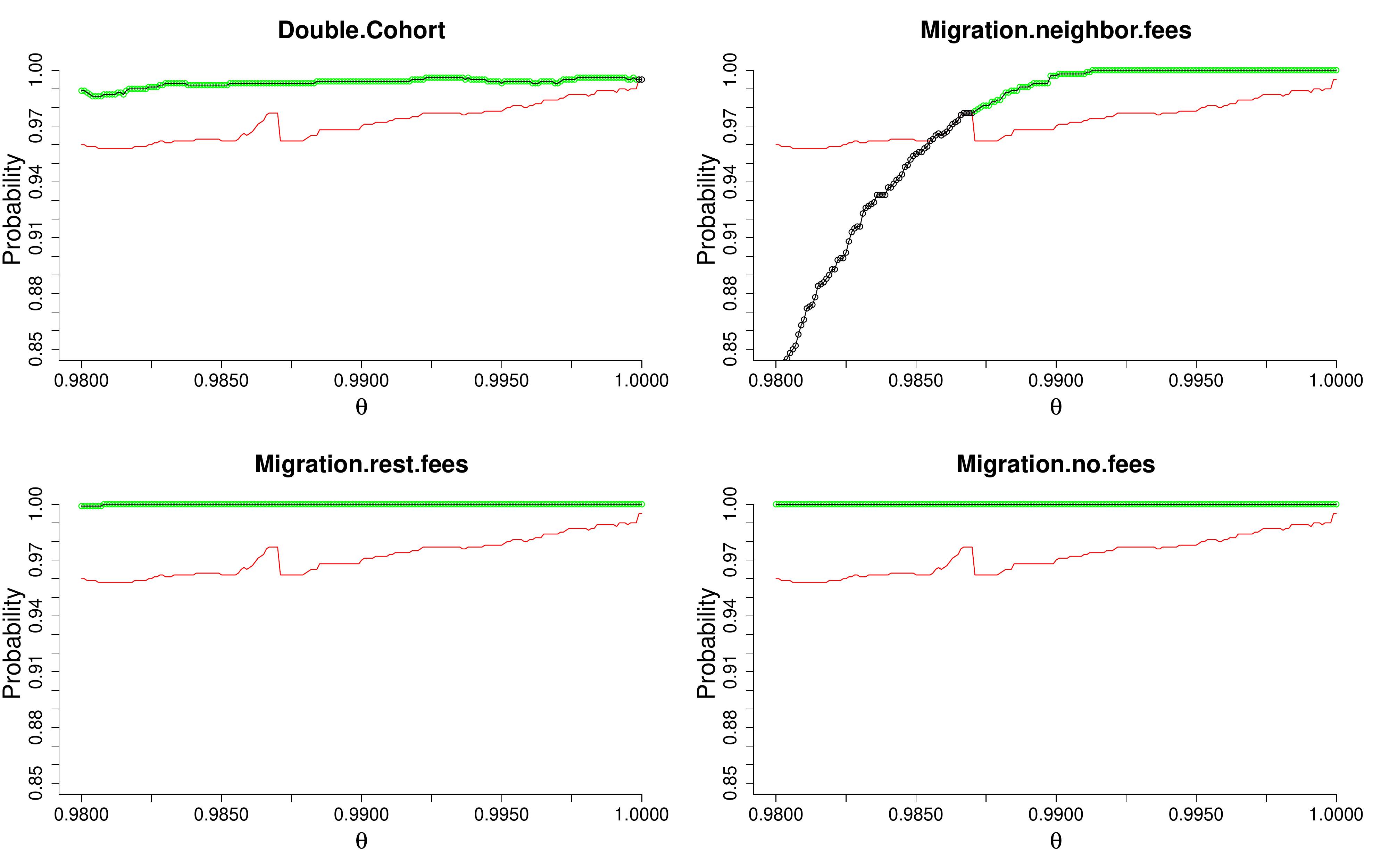}
	
	\caption{Controls with high inclusion probabilities in the first step depending on $\theta$ (x-axis). Green indicates that the respective variable is selected in the final model, which depends on the threshold that is depicted as a red line. The y-axis shows the selection probability from the Lasso step.}
	\label{Fig:controls}
\end{figure}

Figure \ref{Fig:controls} shows all controls that were selected in the main equation \eqref{lasso_outcome} (i.e. with $y_{i,t}$ as the dependent variable). We find the spatial variables to be highly relevant, which implies that mobility and migration effects played a major role for enrollments in the presence of heterogeneous timing and implementation of tuition fees and major educational policy decisions across states.    
In size, they largely contribute in explaining the variability of the enrollment rates. At the lower boundary of $\theta$, only one of the four spatial variables \textit{Migration.neighbor.fees} is included less often over different subsamples and is thus deselected by the stability selection for low $\theta$-values. As there is only a small limited number of overall neighbours of each state, their impact on enrollments in state $i$ is generally much smaller as from the aggregated rest of the country and thus more sensitive to a variation in the response variable.

Furthermore, the variable \textit{Double.cohort} that indicates if there were two cohorts of high school students graduating in the same year, caused by the G8 reform reducing time to graduation, is identified as an important controlling factor. \textit{Double.Cohort} has a negative sign, which at first might appear counter-intuitive, as with a double cohort, one would expect enrollment numbers of students to rise. For relative enrollment rates, however, a negative sign of double cohort seems justified, since universities did not double their admission numbers when there was a double cohort. Moreover, when the competition for universities is extremely high in a double cohort situation, fewer people might decide to actually compete and rather consider outside options or postpone university entrance with a gap year. Note that for the extreme boundary case ($\theta>0.9998$), however, the variable is deselected, which can be attributed to the pre-dominance of the migration factors with large size effects at the extreme upper $\theta$-boundary. Repeating the analysis with \textit{Double.Cohort} in the extreme case for $\theta>0.9998$, however, does not change results and only alters coefficient values in an minor insignificant way. This behavior can be expected when taking into account that the effect of \textit{Double.Cohort} is relatively small compared to the other variables close to the upper boundary of $\theta$.

In line with theory, the variables \textit{Student.to.researcher.ratio} and the share of international enrollments \textit{Migration.international} that are additionally selected in the auxiliary equation of the double selection procedure only have a minor direct influence on enrollment rates, while having a large impact on tuition fees. Thus, this socio-economic factor and the financial situation of universities drives the political decision for the introduction of fees. Overall, the double selection step is key yielding additional necessary variables for accurate estimation of $\beta_0$ (see Figure ~\ref{Fig:grid}).

Generally, these findings show that spatial factors and the double cohort variable are crucial for identifying the effect of tuition fees on enrollments. In the existing empirical literature, however, they have been largely ignored yielding downward biased insignificant estimates. Moreover, the auxiliary equation and the stability double selection are key for detecting the magnitude $\beta_0$.

\subsection{Robustness Checks}
Apart from using all available data, we also analyze two subsets that either contain only periods with tuition fees (2006-2013) or that consist of the peak year 2008 of the presence of tuition fees and the year 2014 after their abolishment. Furthermore, we work with the alternative response variable $y_{i,t}^{\mathit{extra}}=\dfrac{\mathit{NE}_{i,t}}{\mathit{EHG}^*_{i,t}}$ constructed from additional non-public information in the eligible set $\mathit{EHG}^*_{i,t}$ in \eqref{eq:ehgstar}. Estimates of $\beta_{(0)}$ for these adaptations are summarized in Table ~\ref{Tab:Fees_Overview_Robust_design}. Results for HC3 errors do not differ substantially, but are slightly more conservative and can be found in table ~\ref{Tab:Fees_Overview_Robust} in Appendix \ref{Sec:Fig_Tables}.

\begin{table}[]
	\caption{Estimates of the Causal Effect of Tuition Fees $\beta_{(0)}$ for $\theta^*$ in Different Time Frames with Design-Based Standard Errors}
	\label{Tab:Fees_Overview_Robust_design}
	\resizebox{\textwidth}{!}{%
		\begin{tabular}{@{\extracolsep{3pt}}lccccc}
			\hline\hline
			&                                        \multicolumn{3}{c}{Data sets}                                         &  & No. of Variables \\ \cline{2-4}\cline{5-6}
			Tuition Fees                                       &                All                &                Fees                &                Small                &  &  All/Fees/Small  \\ \hline
			&                                   &                                    &                                     &  &                  \\
			\textit{min MSD with} $\theta^*$:                  &         $\mathit{0.9927}$          &          $\mathit{0.9924}$          &          $\mathit{0.9934}$           &  &                  \\
			All Controls                                       &  $\underset{(0.989)}{-1.267 }^{}$  &  $\underset{(1.167)}{-1.952}^{}$   &                  -                  &  &     19/19/-      \\
			Post-Lasso Stability                               &  $\underset{(1.303)}{-2.538}^{}$  &  $\underset{(1.272)}{-2.599}^{*}$  &  $\underset{(1.495)}{-6.345}^{**}$  &  &      4/4/3       \\
			Double Selection Stability                         & $\underset{(1.278)}{-3.996}^{**}$ & $\underset{(1.299)}{-3.180 }^{*}$ & $\underset{(2.488)}{-16.468}^{***}$ &  &      7/6/7       \\
			&                                   &                                    &                                     &  &                  \\
			\textit{min MAD with} $\theta^*$:                  &         $\mathit{0.9927}$          &          $\mathit{0.9926}$          &          $\mathit{0.9945}$           &  &                  \\
			All Controls                                       &  $\underset{(0.989)}{-1.267}^{}$  &  $\underset{(1.168)}{-1.941}^{}$   &                  -                  &  &     19/19/-      \\
			Post-Lasso Stability                               &  $\underset{(1.303)}{-2.538}^{}$  &  $\underset{(1.277)}{-2.599}^{*}$  & $\underset{(1.538)}{-6.126}^{**}$  &  &      4/4/3       \\
			Double Selection Stability                         & $\underset{(1.278)}{-3.996}^{**}$ & $\underset{(1.302)}{-3.185 }^{*}$ & $\underset{(2.549)}{-17.133}^{***}$  &  &      7/6/7       \\
			&                                   &                                    &                                     &  &                  \\
			$y_{i,t}^{\mathit{extra}}$ with $\pi_{1}$/$\pi_2$: & $\mathit{0.999} / \mathit{0.9} $  &  $\mathit{0.9} / \mathit{0.9} $   &   $\mathit{0.85} / \mathit{0.91} $   &  &                  \\
			All Controls                                       &  $\underset{(0.770)}{-1.722}^{*}$  &  $\underset{(0.881)}{-2.213}^{*}$   &                  -                  &  &     19/19/-      \\
			Post-Lasso Stability                               & $\underset{(1.311)}{-3.349}^{*}$  &  $\underset{(0.823)}{-2.234}^{**}$  &   $\underset{(1.317)}{-11.570}^{***}$    &  &  3/9/2      \\
			Double Selection Stability                         & $\underset{(1.087)}{-3.920}^{***}$  & $\underset{(0.877)}{-2.198}^{*}$  &  $\underset{(3.688)}{-15.021}^{**}$  &  &      6/10/6      \\ \hline\hline
		\end{tabular}%
	}
	\caption*{\scriptsize{Note: Response values are scaled to a percentage level. Standard errors in parentheses are calculated based on treatment design and finite populations (\citet{Abadie2020a}). $^{*}$p$<$0.05; $^{**}$p$<$0.01; $^{***}$p$<$0.001 indicate p-values from a t-test on significance from zero. $\theta^*$ is chosen according to minimum mean squared deviation (MSD) and minimum mean absolute deviation (MAD).}}
\end{table}

First, when comparing the effect with $\theta^*$-response values over different time frames, we find that the main results prevail over the variation in the data set. The double selection is still the only reliable method, while post-Lasso and pooled OLS with all controls cannot capture the strength of the effect nor its statistical significance persistently. Post-Lasso generally de-selects too many relevant controls, yielding smaller effects in absolute values of tuition fees on enrollments. Omitting the first and last year from the data only causes mild changes in the amount of included controls, but the size of the estimate for $\beta_0$ from double selection decreases in absolute terms, probably due to fewer available observations. Though, in the extreme case of the smallest data set, where only two years with either ``no fees at all'' or ``fees in seven states'' are considered, the magnitude of the effect increases substantially. The results of the extra response $y_{i,t}^{\mathit{extra}}$ confirm the above observations. The size of the estimates for $\beta_0$ for different time frames and the amount of included controls mostly coincide with results for the response $y_{\theta^*}$. In this case, however, the pure post Lasso double selection estimate is much closer to the estimate of the stability double selection procedure in size and becomes even mildly significant.

In summary, we conclude that the effect is rather robust to changes of the time frame and double selection consistently identifies the effect, where the other methods mostly fail. While changes in the strength of the effect arise mostly in very high-dimensional situations (i.e. small data set), the effect is also identified using the additionally constructed $y_{i,t}^{\mathit{extra}}$. Comparing the strength of the effect to previous studies, which estimated (mostly insignificant) effects from $-0.4$pp to $-2.69$pp, we see that for almost all cases, our estimated effect lies rather between $-3$ and $-4$pp using double selection, and is always highly significant. On the contrary, using fixed effects with all controls and without selection yields estimates that appear to be downwards biased and closer to the lower bound found in other studies, while in almost all cases, this cannot identify significant effects.

\section{Conclusions} \label{Sec:Conclusions}
In this article, we propose a stabilized double selection technique in order to identify the effect of tuition fees on enrollment rates from public state-level data in Germany. We show that such techniques are key for extracting size and significance of the causal effect for the special German situation. In this setting, where few observations coincide with varying implementation and timing of tuition fees and other educational policies across states and time, we are facing correlated covariates and influential observations, which require carefully chosen, tailored econometric techniques.

With our tailored post-Lasso approach, we are the first to find an overall significant negative effect of tuition fees in Germany. With the stability double selection we identify the relevant factors, which are crucial for political decision-making. In particular, previously neglected spatial migration effects and the major shift in educational policy by the G8 high school reform appear as key control variables for enrollment rates in the considered period. The detected effect is robust over a large grid of different response values and different subsets of the full data set. These empirical findings therefore contribute to the existing literature on education economics. In the active ongoing discussion about the reintroduction of tuition fees in Germany, the results might also be of political interest.

Moreover, this study strongly advocates the use of data-driven variable selection to choose relevant controls from a broad set of possible influencing factors. We explicitly show that standard fixed effects panel regressions without selecting variables fails to detect correct and precise effects for such small sample sizes relative to the dimensionality of the problem. Furthermore, appropriate statistical selection techniques determine and justify the relevance of chosen controlling factors, yielding an easily interpretable post-selection model that outperforms all ad-hoc choices.
For future research, it would be interesting to use the data-driven identification of relevant controls also for other countries, e.g. the United Kingdom or France, aiming for a comprehensive European study with increasingly relevant spatial cross-effects across country borders. This is particularly relevant given the reintroduction of fees for international students in parts of Germany, that could trigger such cross-effects.

%\section{References}
%\newpage
\bibliographystyle{apalike}
\bibliography{literature}

\newpage

\appendix
\section{Appendix}
\subsection{Algorithm for Threshold Choice}
\label{Sec:Appendix_Threshold}
To obtain the thresholds for different $\theta$-values in the full data set, we compute the thresholds automatically using the following algorithm:
\begin{enumerate}
\item For a given $\tilde{\theta}$, order the inclusion frequencies $\Pi_{\tilde{j}}^1 \in \{\Pi_{j}^1, \ j=1,\dots,p \}$, $\tilde{j}=1,\dots,p$ and obtain $\Pi_{\tilde{1}}^1,\dots,\Pi_{\tilde{j}}^1,\dots,\Pi_{\tilde{p}}^1$, i.e. $\Pi_{\tilde{1}}^1\geq \dots\geq \Pi_{\tilde{j}}^1\geq\dots\geq \Pi_{\tilde{p}}^1$.
\item Compute the difference $\Delta_j=\Pi_{\tilde{j}}^1-\Pi_{\tilde{j}+1}^1$ for all $A=\{\tilde{j}: \Pi_{\tilde{j}+1}^1>\pi_{\mathit{min}}\}$, i.e. look at the distance between inclusion frequencies.
\item Choose the cutoff $\pi_{1,\tilde{\theta}}$ to lie at index $\hat{j}=\max \big\{\underset{\tilde{j}\in A}{argmax}\Delta_{\tilde{j}}\big\}$, and obtain $\pi_{1,\tilde{\theta}}=\Pi_{\hat{j}+1,1}$.
\end{enumerate}
The algorithm chooses the cutoff at a large difference between ordered inclusion frequencies, i.e. where noise variables are distinguished by true influencing variables. The minimum threshold  $\pi_{1,\theta}^{\mathit{min}}$ is set to ensure that the largest difference between inclusion frequencies does not occur between two noise variables (i.e. with very low inclusion frequency). We set this $\pi_{1,\theta}^{\mathit{min}}, \ \theta \in [0.98,1]$ in the following way to make sure noise variables are screened out:

\[
\pi_{1,\theta}^{\mathit{min}}=
\begin{cases}
0.945 \  & \text{\textit{All} Data set and } \theta\in[0.98,0.992] \\
0.975 \  & \text{\textit{All} Data set and } \theta\in(0.992,1] \\
0.98 \  & \text{\textit{Fees} Data set and }  \theta \in [0.98,1] \\
0.9 \  & \text{\textit{Small} Data set and }  \theta \in [0.98,1] \\
\end{cases} \ .
\]
For the large data set, the minimum threshold is adapted once as $\bar{y}$-values rise strongly with $\theta$, especially when $\theta$ is close to 1. There, all variables are have higher inclusion frequencies which makes it necessary to adapt $\pi_{1,\theta}^{\mathit{min}}$.\\

For $\pi_2$, we have no variation in $\theta$ and compute the thresholds in the same manner as above manually. We obtain
\[
\pi_{2}=
\begin{cases}
0.9 \  & \text{\textit{All} Data set} \\
0.9 \  & \text{\textit{Fees} Data set} \\
0.93 \  & \text{\textit{Small} Data set} \\
\end{cases} \ .
\]

\subsection{Design-Based Standard Errors}\label{sec:appendix_se}
In the following, we describe the detailed step-wise version of obtaining the design-based standard errors in \eqref{eq:se} as proposed in \citet{Abadie2020a}:
\begin{enumerate}
	\item Let $U=(X \quad Z)$ and calculate $V=D - \Lambda U$, where $\Lambda=\Big(U\transpose U\Big)^{-1} U\transpose D$ is the least squares estimator of regressing $D$ on $U$.
	\item Calculate the residuals $\hat{\ddot{\epsilon}}^{(1)}_{i,t}$ from the least-squares regression in equation \ref{eq:postl} and obtain $\hat{V}_{\hat{\ddot{\epsilon}}}=\Big(V_{1,1}\hat{\ddot{\epsilon}}^{(1)}_{1,1},\dots,V_{i,t}\hat{\ddot{\epsilon}}^{(1)}_{i,t},\dots,V_{n,T}\hat{\ddot{\epsilon}}^{(1)}_{n,T}\Big)\transpose$.
	\item Calculate the least-squares estimator $\hat{\beta}_{\epsilon}$ of a regression of $\hat{V}_{\hat{\ddot{\epsilon}}}$ on $U$ as $\hat{\beta}_{\epsilon}=\Big(U\transpose U\Big)^{-1}U\transpose \hat{V}_{\hat{\ddot{\epsilon}}}$. The fitted values of this regression serve as an estimate of $E[\hat{V}_{\hat{\ddot{\epsilon}}}]$, which is needed to calculate $Var[\hat{V}_{\hat{\ddot{\epsilon}}}]$.
	\item Calculate $G=Var[\hat{V}_{\hat{\ddot{\epsilon}}}]=\Big(\hat{V}_{\hat{\ddot{\epsilon}}}-U\hat{\beta}_{\epsilon}\Big)\transpose \Big(\hat{V}_{\hat{\ddot{\epsilon}}}-U\hat{\beta}_{\epsilon}\Big)$
	\item Calculate $SE(\beta_{(0)})=\sqrt{(V\transpose V)^{-1}G(V\transpose V)^{-1}}$.
\end{enumerate}

\subsection{Data}
\subsubsection{Individual SOEP Data}
\label{Sec:Appendix_SOEP_1}
\begin{figure}[!htb]
	\centering
	\includegraphics[width=\textwidth]{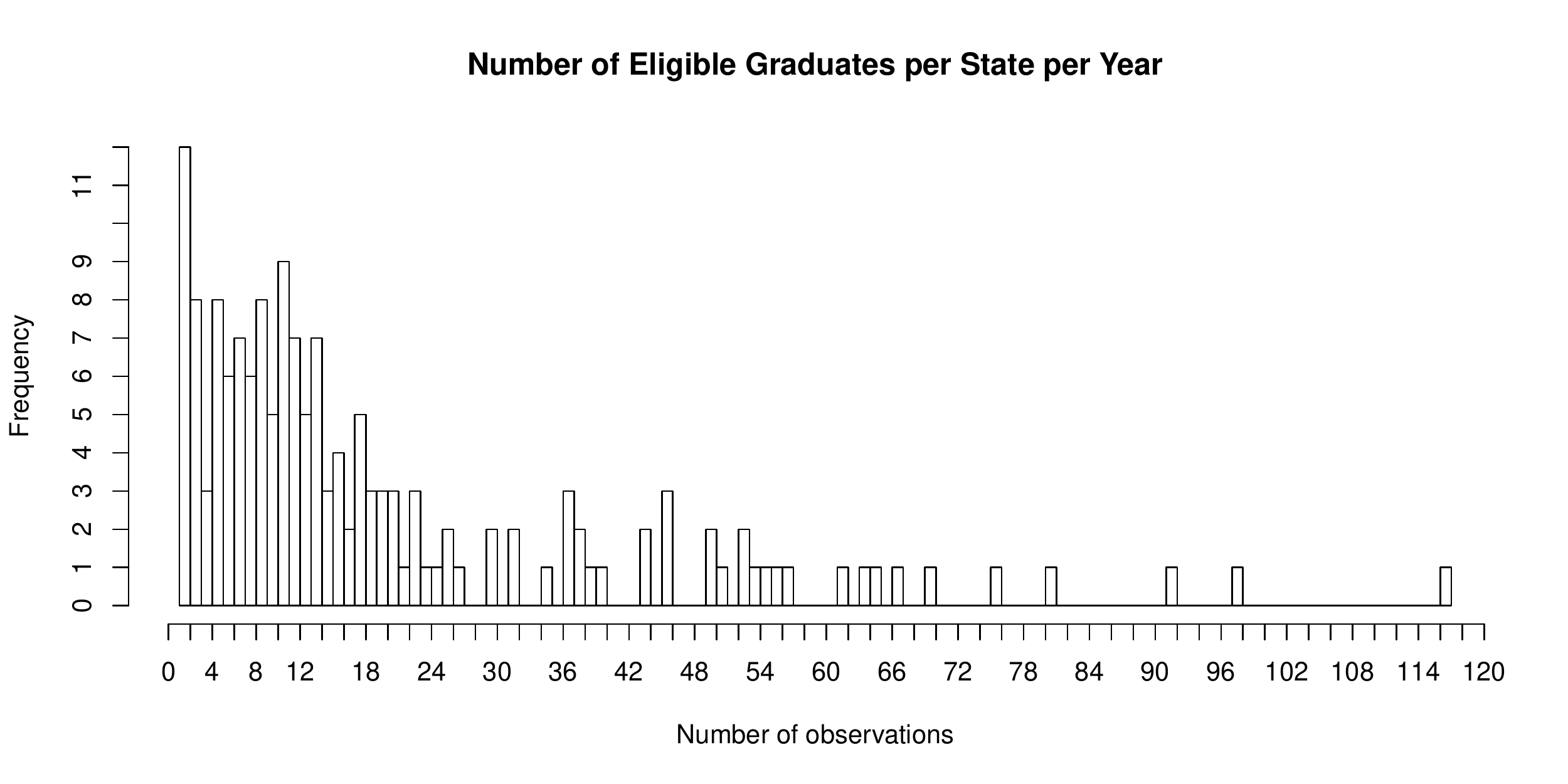}
	\caption{Histogram of the number of eligible high school graduates in each state and year (i.e. 160 tuples) in the SOEP data set \textit{edubio}. There was one state-year combination with no observations at all.}
	\label{fig:soep_eligible}
\end{figure}

\begin{figure}[!htb]
	\centering
	\includegraphics[width=\textwidth]{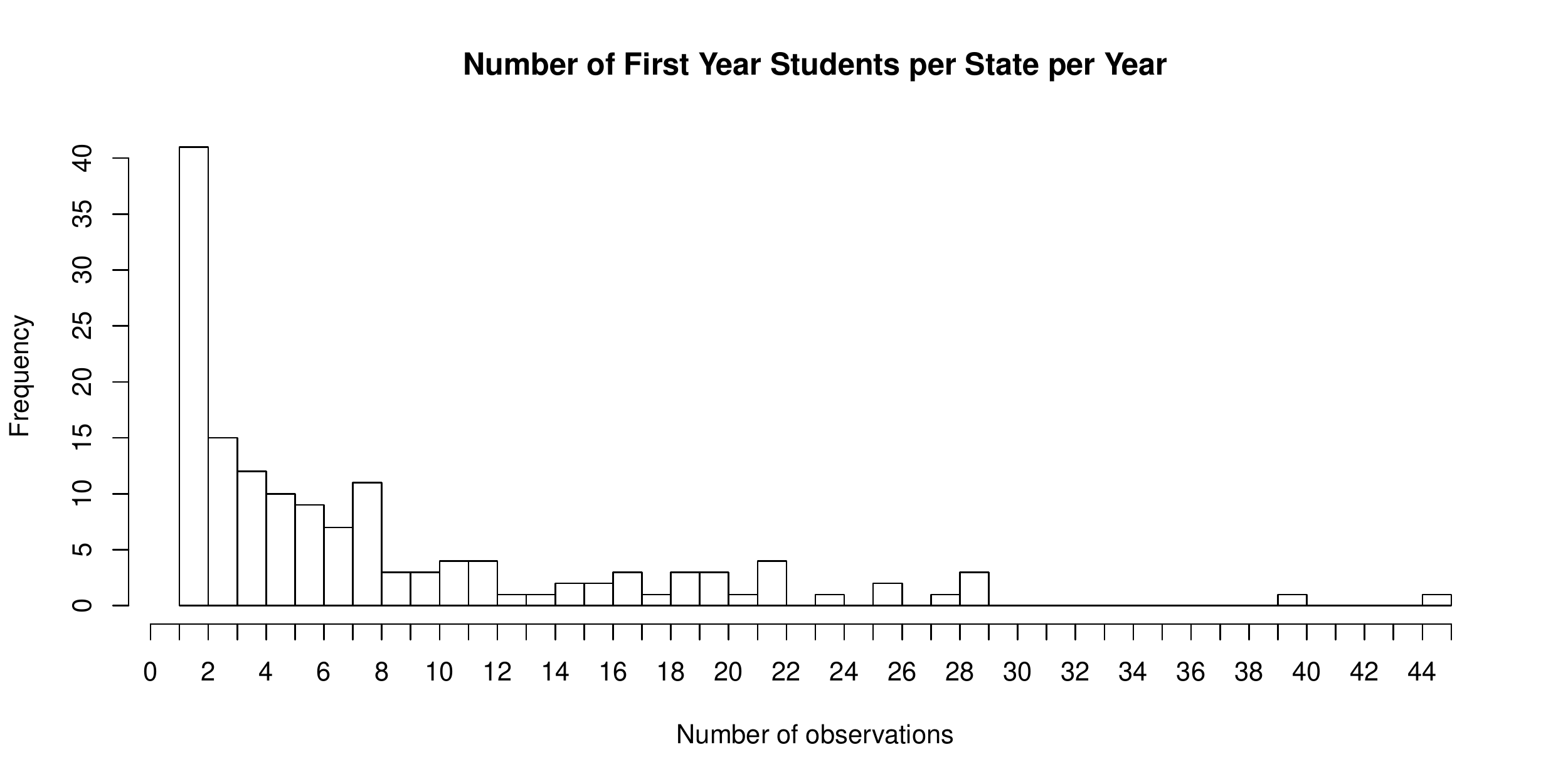}
	\caption{Histogram of the number of first year students in each state and year (i.e. 160 tuples) in the SOEP data set \textit{edubio}. There were $11$ state-year combinations with no observations at all.}
	\label{fig:soep_students}
\end{figure}

Figures \ref{fig:soep_eligible} and \ref{fig:soep_students} give an overview on the lack of observations of individuals in the SOEP data set. On the $x$-axis, the number of observations for each state-year tuple is shown, whereas on the $y$-axis, the frequency of tuples with the specific number of observations is depicted. As can be seen in the histograms, there were many tuples with insufficient number of observations to represent a state. More specifically, 109 tuples out of 160 have less than 20 observations each (i.e. individuals) in Figure \ref{fig:soep_eligible} (i.e. eligible high school graduates in state $i$ and year $t$) and 119 tuples out of 160 have less than 10 observations each (i.e. individuals) in Figure \ref{fig:soep_students} (i.e. first year students in state $i$ and year $t$). This makes it necessary to use publicly available data aggregated on a state-level instead of individual data, since the latter cannot be representative for a state's specific cohort (i.e. taking less than 20 observations to represent an entire cohort for the majority of tuples).
\subsubsection{Information on Control Variables from the SOEP}
The control variables can be split up into socio-economic variables (Table \ref{Tab:Descriptive_socio}), variables describing university statistics (Table \ref{Tab:Descriptive_Uni_Students}) and spatial variables (Table \ref{Tab:Spatial_Controls}).\\
Even though the number of households can vary, this cannot bias results since we measure shares of the population. In some cases, there is a substantial amount of missing values (i.e. "Rent", "Income"), which does not pose a large problem as enough data points are still available. These pre-chosen variables are all publicly available and are all potentially correlated with the outcome or the effect.\\
For the Destatis-data, the variables are already aggregated on a state level, while for the SOEP-data, the aggregation is done manually using the HID\footnote{HID stands for Household-ID. For the variable "Life Satisfaction", data was available on a personal level. This does not make a difference since mean values are used.}, which identifies a household over different subsamples and over time. All SOEP variables are therefore mean values (rent, income, life satisfaction).\\
\newpage
\section{Figures and Tables}
\label{Sec:Fig_Tables}
\begin{figure}[!htb]
	\centering
	\includegraphics[width=\textwidth]{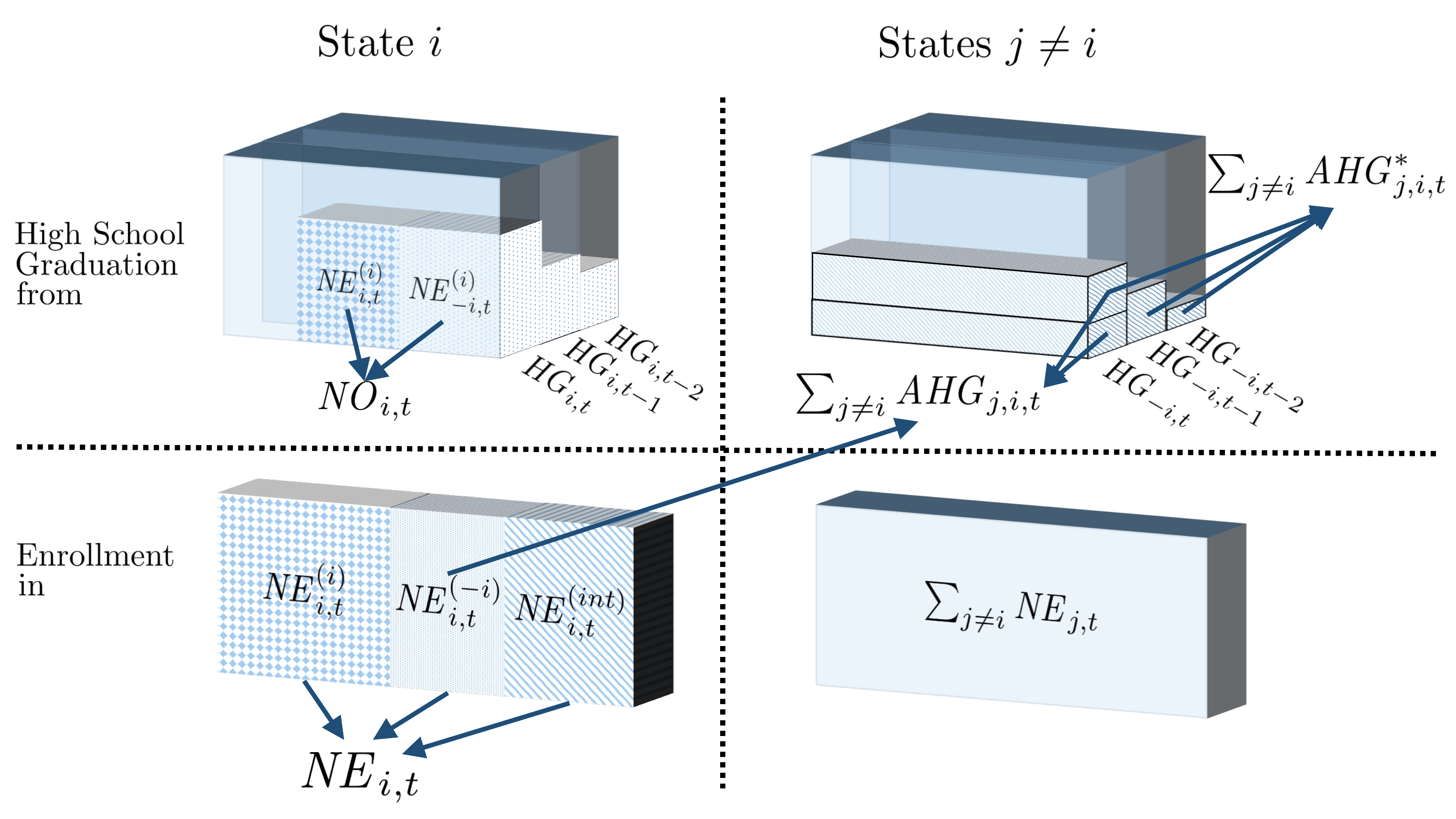}
	\caption{Illustration of the composition of the response variable. High school graduates from different cohorts are depicted in the top boxes, new enrollments in the bottom boxes. The index $-i$ describes all states $j=1,\dots,16$ except the $i$th state.}
	\label{fig:var_response}
\end{figure}

\begin{table}[]
	\caption{Summary of Defined Quantities for the Response }
	%Illustration of $\mathit{NO}_{j,t}$ and $\mathit{NE}_{i,t}$
	\label{Tab:Var_expl}
	%\resizebox{\textwidth}{!}{%
	\begin{tabular}{@{\extracolsep{6pt}}p{5.5cm}l}
		\hline\hline
		Enrollments in state $i$ from anywhere:                                                                                                                                                                                                                               & $\displaystyle\mathit{NE}_{i,t}=\sum_{j=1}^{n}\mathit{NE}^{(j)}_{i,t}+\mathit{NE}^{(int)}_{i,t}=\sum_{j=1}^{n}\Big(\sum_{\tau=t-\psi}^{t}\mathit{NE}^{(j)}_{\tau,i,t}\Big) +\mathit{NE}^{(int)}_{i,t}$ \\
		Enrollments from one state $j$ to anywhere in Germany:                                                                                                                                                                                                                           & $\displaystyle\mathit{NO}_{j,t}=\sum_{\tau=t-\psi}^{t}\mathit{NO}_{\tau,j,t}=\sum_{\tau=t-\psi}^{t}\sum_{i=1}^{n} \mathit{NE}^{(j)}_{\tau,i,t}$ \\
		Eligible set of high school graduates in $i$: & $\mathit{EHG}_{i,t}=\mathit{NE}^{(int)}_{i,t}+ \mathit{HG}_{i,t} +\sum_{j\neq i}\mathit{AHG}_{j,i,t}$  \\
		International Enrollments to $i$: & $\displaystyle\mathit{NE}^{(int)}_{i,t}$ \\
		High school graduates from $i$: & $\displaystyle \mathit{HG}_{i,t}$ \\
		High school graduates from $j$ affected by enrollments in $i$:& $\displaystyle\mathit{AHG}_{j,i,t}=\theta NE_{i,t}^{(j)} + (1-\theta)\mathit{HG}_{j,t}$\\
		\hline\hline
	\end{tabular}%
	\caption*{\scriptsize{Note: $\psi$ stands for the maximum number of years after which a high school graduate enrolled to a german university. Here, the index $\tau$ or rather $\psi$ is set so that it includes all students from earlier cohorts.}}
\end{table}

\begin{figure}[h]
	\centering
	\includegraphics[width=\textwidth]{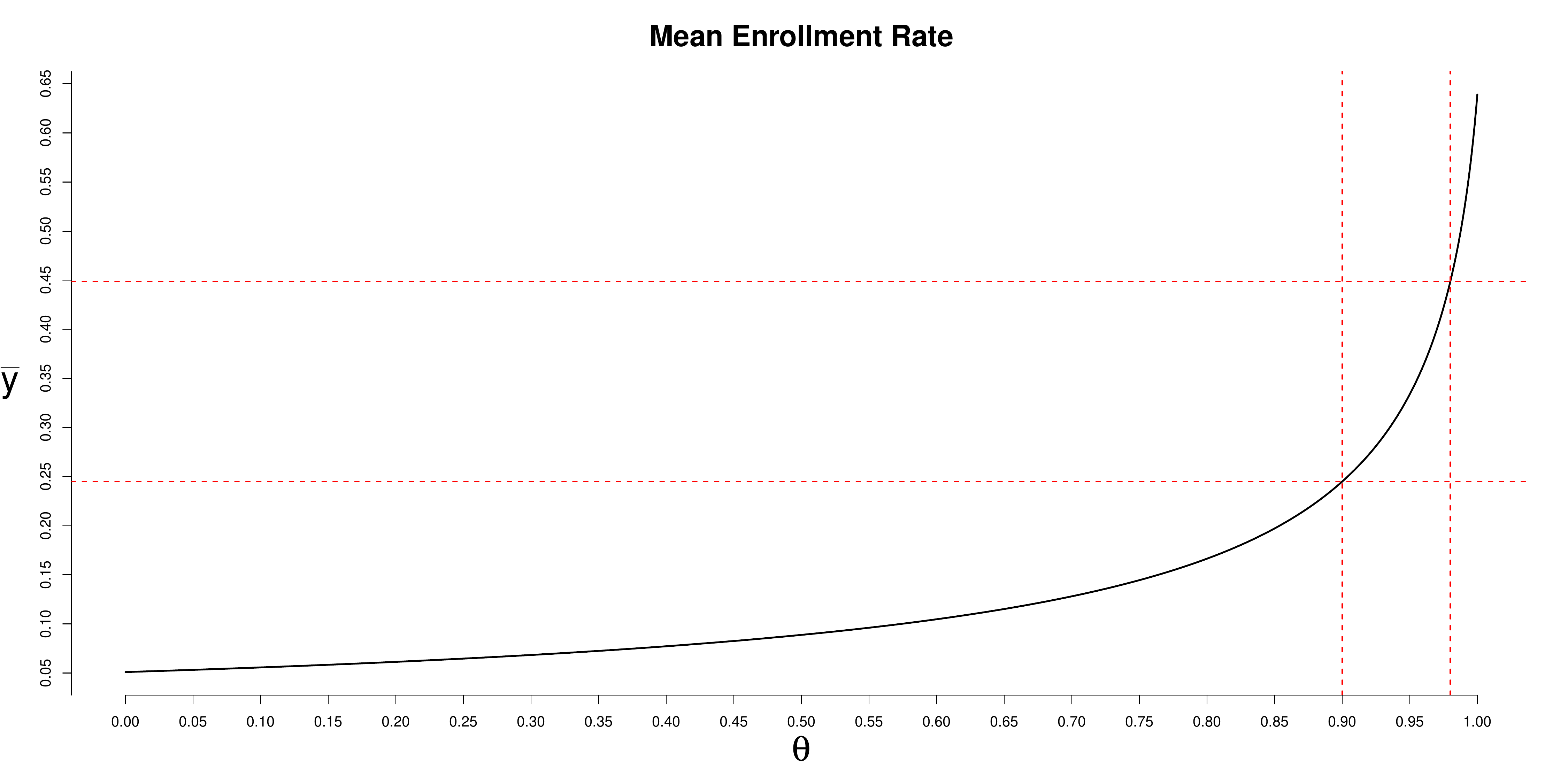}
	\caption{Illustration of mean values of the response variable for a given value of $\theta$. The red dotted lines show the lower boundaries $\theta=0.9$ and $\theta=0.98$ with their respective mean response value.}
	\label{fig:y_mean}
\end{figure}
\clearpage

\begin{minipage}{\textwidth}
	\captionof{table}{Description of University and Student Control Variables}
			\resizebox{1\textwidth}{!}{%
			\begin{tabular}{lp{0.5\textwidth}p{0.3\textwidth}}
				\hline\hline
				Variable                         & Description                                                                                                                                                                                                                                                               &   Source                                                \\ \hline
				&                                                                                                                                                                                                                                                                           &                                                      \\
				\multicolumn{2}{l}{\textbf{Control Variables regarding Students and Universities} }  &                                                                                                                                                                                                                                                                                                                                    \\ \\
				Log.Third.Party.Funds.per.institution                 & Natural Logarithm of the quotient of third party funds for universities in state $i$ in year $t$ divided by the number of state accredited higher institutions, aggregated at a state level                                                                                                                                                                      & \citet{FederalStatisticalOffice_Monetaer}: 2.1.3; \url{www.hochschulkompass.de}     \\
				Log.Spendings.per.Student                  & Natural Logarithm of spendings of state $i$ in year $t$ per student, aggregated over all universities $i$                                                                                                                                                                                  & \citet{FederalStatisticalOffice_Monetaer}: 1.1    \\
				Student.to.researcher.ratio                        & Number of students in state $i$ in year $t$ per scientific employee of higher institutions                                                                                                   & \citet{FederalStatisticalOffice_Personal}: ZUS-01                                        \\
				Habilitations                    & Share of habilitations at universities in state $i$ to habilitations over all states                                                                                                                                                                                        & \citet{FederalStatisticalOffice_Personal}: ZUS-07     \\
				Graduates                        & Share of graduates at universities in state $i$ to relevant population                                                                                                                                                                                               & \citet{FederalStatisticalOffice_Nichtmonetaer}: TAB-02                                        \\
				Women.Studying                   & Share of female students studying at higher institutions to all students                                                                                                                                                                                                    & Genesis: Table 21311-0014                             \\ \hline
			\end{tabular}%
		}
		\captionof*{table}{\scriptsize{Note: data from the Federal Statistical Office are availible in the ".xls" format in German in the respective sheet (indicated by TAB or ZUS). They can be found at the bottom of the page \url{https://www.destatis.de/DE/ZahlenFakten/GesellschaftStaat/BildungForschungKultur/Hochschulen/Hochschulen.html} under "Ausgewählte Publikationen", using the name of the report and the reference. Reports from earlier years are in the report of the respective year. For the Genesis Data, the tables are found at \url{https://www-genesis.destatis.de} in the menu under "Available Data" on the page "Tables". There, a search for a specific coding leads to the desired tables. The data is gathered by choosing the according year.}
		}
		\label{Tab:Descriptive_Uni_Students}

\end{minipage}

\begin{table}[]
	\caption{Description of Spatial Control Variables}
	\resizebox{\textwidth}{!}{%
		\begin{tabular}{lp{0.6\textwidth}p{0.3\textwidth}}
			\hline\hline
			Variable                         & Description                                                                                                                                                                                                                                                               &   Source                                                \\ \hline
			&                                                                                                                                                                                                                                                                           &                                                      \\
			
			\multicolumn{2}{l}{\textbf{Spatial Control Variables} }  &                                                                                                                                                                                                                                                                                                                                    \\ \\
			Migration.neighbor.fees                        & Enrolling students to state $i$ in year $t$ with high school diploma from fee neighbor states (i.e. states that share a border with $i$ and that have tuition fees in the winter term of year $t$) minus enrollments to fee neighbor states by students with high school diploma from state $i$, both divided by all new enrollments in state $i$.                                                                                                                                                    & \citet{FederalStatisticalOffice_Nichtmonetaer}: TAB-13                                        \\
			Migration.rest.fees                        & Enrolling students to state $i$ in year $t$ with high school diploma from fee non-neighbor states (i.e. states that do not share a border with $i$ and that have tuition fees in the winter term of year $t$) minus enrollments to fee non-neighbor states by students with high school diploma from state $i$, both divided by all new enrollments in state $i$.                                                                                                                                                 & \citet{FederalStatisticalOffice_Nichtmonetaer}: TAB-13                                        \\
			Migration.no.fees                        & Enrolling students to state $i$ in year $t$ with high school diploma from non-fee states (i.e. states that do not have tuition fees in the winter term of year $t$) minus enrollments to non-fee states by students with high school diploma from state $i$, all divided by all new enrollments in state $i$.                                                                                                                                                    & \citet{FederalStatisticalOffice_Nichtmonetaer}: TAB-13                                        \\
			Migration.international                        & Share of new enrollments of international students (i.e. students that did not obtain their high school diploma in Germany) to state $i$ in year $t$ relative to all new enrollments in state $i$                                                                                                                                              & \citet{FederalStatisticalOffice_Nichtmonetaer}: TAB-13                                        \\ \hline
		\end{tabular}%
	}
	\caption*{\scriptsize{Note: data from the Federal Statistical Office are availible in the ".xls" format in German in the respective sheet (indicated by TAB or ZUS). They can be found at the bottom of the page \url{https://www.destatis.de/DE/ZahlenFakten/GesellschaftStaat/BildungForschungKultur/Hochschulen/Hochschulen.html} under "Ausgewählte Publikationen", using the name of the report and the reference. Reports from earlier years are in the report of the respective year. More details on the calculation and choice of spatial variables in text.}
	}
	\label{Tab:Spatial_Controls}
\end{table}
\begin{table}[]
	\caption{Description of Regression Variables and Socio-Economic Control Variables}
	\resizebox{\textwidth}{!}{%
		\begin{tabular}{p{0.35\textwidth}p{0.62\textwidth}p{0.3\textwidth}}
			\hline\hline
			Variable                            & Description                                                                                                                                                                                                                                        & Source                                                                                                                               \\ \hline
			                                    &                                                                                                                                                                                                                                                    &                                                                                                                                      \\
			\textbf{Regression Variables}       &                                                                                                                                                                                                                                                    &                                                                                                                                      \\
			$y_{i,t}$                     & Enrollment rate: new first year students in state $i$ and winter term of year $t/t+1$ divided by affected graduates. Affected graduates consist of high school graduates in state $i$ and year $t$, international first year students in state $i$ and winter term of year $t/t+1$ and a weighted number of in-country migration from other German states ($\sum_{j\neq i} \mathit{EHG}_{j,i,t}$). $\mathit{EHG}_{j,i,t}$ is calculated using convex combinations of $\mathit{NE}^{(j)}_{i,t}$ and $\mathit{HG}_{j,t}$.  & Own calculation from \citet{FederalStatisticalOffice_Nichtmonetaer}: TAB-13 and \citet{FederalStatisticalOffice_Studierende}: TAB-06 \\
			$y_{i,t}^{\mathit{extra}}$                     &   Similar to $y_{i,t}$, but using a different measurement for $\mathit{EHG}_{j,i,t}$ (see Section \ref{Sec:Data}).        & Own calculation from \citet{FederalStatisticalOffice_Nichtmonetaer}: TAB-13 and \citet{FederalStatisticalOffice_Studierende}: TAB-06 \\
			$d_{i,t}$: Tuition.Fees                           & 1 if tuition fees were present in state $i$ in winter term of year $t/t+1$, zero otherwise                                                                                                                                                         & \citet{Mitze2015}                                                                                                                    \\
			\textbf{Socio-Economic Statistics}  &                                                                                                                                                                                                                                                    &                                                                                                                                      \\
			log.Rent                            & Natural logarithm of average rent in households of state $i$ in year $t$ excluding heating or extra costs                                                                                                                                          & SOEP: Hgen, Hgrent                                                                                                                   \\
			log.Income                          & Natural logarithm of average income in households of state $i$ in year $t$                                                                                                                                                                         & SOEP :Hgen, Hghinc                                                                                                                   \\
			Urbanization.level                  & Share of households living in cities in state $i$ in year $t$                                                                                                                                                                                      & SOEP: Hbrutto, Regtyp                                                                                                                \\
			Life.Satisfaction                   & Average life satisfaction per person (0=Completely dissatisfied, 1= Completely satisfied) in state $i$                                                                                                                                             & SOEP: pequiv, P11101                                                                                                                 \\
			Unemployment.Rate                   & Unemployment rate in state $i$ (0=0\%, 1=100\%)                                                                                                                                                                                                    & Genesis                                                                                                                              \\
			G8                                  & 1 if students graduated high school in 8 years in state $i$ in year $t$, zero otherwise                                                                                                                                                            & Own research                                                                                                                         \\
			Double.Cohort                       & 1 if there was a double cohort of students graduating high school in state $i$ in year $t$, zero otherwise                                                                                                                                         & Own research                                                                                                                         \\
			Mil.Service                    & 1 if there was mandatory military service for male high school graduates in Germany in year $t$, zero otherwise                                                                                                                                    & Own research                                                                                                                         \\ \hline
		\end{tabular}%
	}
	\caption*{\scriptsize{All data sets used to obtain the variables can be found in the SOEP-database at \url{https://www.diw.de/en/soep} with the corresponding variable description and sample (SOEP: sample, variable) at \url{https://paneldata.org/soep-long}.   For the Genesis Data, the tables are found at \url{https://www-genesis.destatis.de} in the menu under "Available Data" on the page "Tables". There, a search for a specific coding leads to the desired tables. The data is gathered by choosing the respective year.}
	}
	\label{Tab:Descriptive_socio}
\end{table}

\afterpage{
	\clearpage
	\begin{landscape}
		\begin{table}[!htbp] \centering
			\captionof{table}{Simulation Results for Different Forms and Strengths of Influential Observations with HC3 Standard Errors}
			\label{Tab:Sim_results}
			\resizebox{1.35\textheight}{!}{%
				\begin{tabular}{
						%lp{0.1\textwidth}cp{0.1\textwidth}cp{0.1\textwidth}cp{0.1\textwidth}cp{0.1\textwidth}cp{0.1\textwidth}c}
						l@{\hskip 20pt}ccc@{\hskip 30pt} ccc@{\hskip 30pt} ccc@{\hskip 30pt} ccc@{\hskip 30pt} ccc@{\hskip 30pt} ccc@{\hskip 30pt}}
					\\[-1.8ex]\hline
					\hline \\[-1.8ex] & \multicolumn{3}{c@{\hskip 30pt}}{Absolute $\mathit{Bias}_{\eta_0}$} & \multicolumn{3}{c@{\hskip 30pt}}{$\mathit{RMSE}_{\eta_0}$} & \multicolumn{3}{c@{\hskip 30pt}}{$\#$ Covariates} & \multicolumn{3}{c@{\hskip 30pt}}{TPR} & \multicolumn{3}{c@{\hskip 30pt}}{FPR} & \multicolumn{3}{c@{\hskip 30pt}}{Rejection Rate}   \\
					\cmidrule(r{30pt}){2-4} \cmidrule(r{30pt}){5-7} \cmidrule(r{30pt}){8-10} \cmidrule(r{30pt}){11-13} \cmidrule(r{30pt}){14-16} \cmidrule(r{30pt}){17-19} \\[-1.8ex]
					\textit{Size of Distortion:} & 0 & 1 & 5 & 0 & 1 & 5 & 0 & 1 & 5 & 0 & 1 & 5 & 0 & 1 & 5 & 0 & 1 & 5\\ \midrule \\[-1.8ex]
					\multicolumn{3}{l}{\textit{Distortion in the Active Set}}  \\
					PL Stab: 0.5 & 0.156 & 0.158 & 0.181 & 0.025 & 0.026 & 0.034 & 9.238 & 9.338 & 7.454 & 0.837 & 0.839 & 0.734 & 0.043 & 0.047 & 0.006 & 0.999 & 0.994 & 1.000 \\
					DB Stab: 0.5 & 0.088 & 0.087 & 0.081 & 0.024 & 0.024 & 0.021 & 10.669 & 10.629 & 10.099 & 1.000 & 1.000 & 1.000 & 0.034 & 0.032 & 0.005 & 0.068 & 0.062 & 0.058 \\
					PL Stab: 0.7 & 0.161 & 0.164 & 0.192 & 0.027 & 0.028 & 0.038 & 8.360 & 8.344 & 6.808 & 0.802 & 0.801 & 0.679 & 0.017 & 0.017 & 0.001 & 1.000 & 1.000 & 1.000 \\
					DB Stab: 0.7 & 0.087 & 0.086 & 0.081 & 0.024 & 0.023 & 0.021 & 10.190 & 10.199 & 10.021 & 1.000 & 1.000 & 1.000 & 0.009 & 0.010 & 0.001 & 0.066 & 0.059 & 0.056 \\
					Post Lasso & 0.143 & 0.143 & 0.131 & 0.024 & 0.024 & 0.025 & 19.395 & 19.156 & 14.289 & 0.917 & 0.917 & 0.912 & 0.511 & 0.499 & 0.259 & 0.842 & 0.839 & 0.673 \\
					Double Selection & 0.090 & 0.092 & 0.086 & 0.025 & 0.026 & 0.023 & 21.230 & 20.703 & 16.395 & 1.000 & 1.000 & 1.000 & 0.561 & 0.535 & 0.320 & 0.065 & 0.065 & 0.061 \\
					Fixed Effects All & 0.092 & 0.094 & 0.089 & 0.028 & 0.028 & 0.026 & 30.000 & 30.000 & 30.000 & 1.000 & 1.000 & 1.000 & 1.000 & 1.000 & 1.000 & 0.051 & 0.058 & 0.045 \\
					Oracle & 0.086 & 0.086 & 0.081 & 0.024 & 0.023 & 0.021 & 10.000 & 10.000 & 10.000 & 1.000 & 1.000 & 1.000 & 0.000 & 0.000 & 0.000 & 0.068 & 0.060 & 0.056 \\
					\\[-1.8ex]
					\multicolumn{3}{l}{\textit{Distortion in the Inactive Set}}  \\
					PL Stab: 0.5 & 0.156 & 0.157 & 0.157 & 0.025 & 0.026 & 0.026 & 9.238 & 9.435 & 9.221 & 0.837 & 0.840 & 0.842 & 0.043 & 0.052 & 0.040 & 0.999 & 0.994 & 0.996 \\
					DB Stab: 0.5 & 0.088 & 0.087 & 0.087 & 0.024 & 0.024 & 0.024 & 10.669 & 10.750 & 10.585 & 1.000 & 1.000 & 1.000 & 0.034 & 0.037 & 0.029 & 0.068 & 0.066 & 0.066 \\
					PL Stab: 0.7 & 0.161 & 0.164 & 0.162 & 0.027 & 0.028 & 0.027 & 8.360 & 8.354 & 8.306 & 0.802 & 0.801 & 0.802 & 0.017 & 0.017 & 0.014 & 1.000 & 1.000 & 1.000 \\
					DB Stab: 0.7 & 0.087 & 0.086 & 0.086 & 0.024 & 0.023 & 0.023 & 10.190 & 10.251 & 10.153 & 1.000 & 1.000 & 1.000 & 0.009 & 0.013 & 0.008 & 0.066 & 0.061 & 0.061 \\
					Post Lasso & 0.143 & 0.144 & 0.144 & 0.024 & 0.025 & 0.025 & 19.395 & 19.275 & 18.535 & 0.917 & 0.918 & 0.918 & 0.511 & 0.505 & 0.468 & 0.842 & 0.834 & 0.840 \\
					Double Selection & 0.090 & 0.092 & 0.091 & 0.025 & 0.026 & 0.026 & 21.230 & 21.122 & 20.548 & 1.000 & 1.000 & 1.000 & 0.561 & 0.556 & 0.527 & 0.065 & 0.067 & 0.072 \\
					Fixed Effects All & 0.092 & 0.094 & 0.094 & 0.028 & 0.028 & 0.028 & 30.000 & 30.000 & 30.000 & 1.000 & 1.000 & 1.000 & 1.000 & 1.000 & 1.000 & 0.051 & 0.062 & 0.061 \\
					Oracle & 0.086 & 0.087 & 0.087 & 0.024 & 0.023 & 0.023 & 10.000 & 10.000 & 10.000 & 1.000 & 1.000 & 1.000 & 0.000 & 0.000 & 0.000 & 0.068 & 0.061 & 0.061 \\
					\\[-1.8ex]
					\multicolumn{3}{l}{\textit{Distortion in the Response}}   \\
					PL Stab: 0.5 & 0.156 & 0.158 & 0.169 & 0.025 & 0.026 & 0.030 & 9.238 & 9.306 & 8.537 & 0.837 & 0.836 & 0.769 & 0.043 & 0.047 & 0.042 & 0.999 & 0.998 & 0.999 \\
					DB Stab: 0.5 & 0.088 & 0.088 & 0.107 & 0.024 & 0.024 & 0.036 & 10.669 & 10.788 & 10.777 & 1.000 & 1.000 & 1.000 & 0.034 & 0.039 & 0.039 & 0.068 & 0.062 & 0.065 \\
					PL Stab: 0.7 & 0.161 & 0.164 & 0.176 & 0.027 & 0.028 & 0.032 & 8.360 & 8.330 & 7.550 & 0.802 & 0.797 & 0.726 & 0.017 & 0.018 & 0.015 & 1.000 & 1.000 & 1.000 \\
					DB Stab: 0.7 & 0.087 & 0.088 & 0.106 & 0.024 & 0.024 & 0.035 & 10.190 & 10.237 & 10.238 & 1.000 & 1.000 & 1.000 & 0.009 & 0.012 & 0.012 & 0.066 & 0.063 & 0.054 \\
					Post Lasso & 0.143 & 0.143 & 0.149 & 0.024 & 0.025 & 0.029 & 19.395 & 19.361 & 19.019 & 0.917 & 0.917 & 0.908 & 0.511 & 0.509 & 0.497 & 0.842 & 0.834 & 0.817 \\
					Double Selection & 0.090 & 0.092 & 0.111 & 0.025 & 0.026 & 0.039 & 21.230 & 21.153 & 21.392 & 1.000 & 1.000 & 1.000 & 0.561 & 0.558 & 0.570 & 0.065 & 0.069 & 0.053 \\
					Fixed Effects All & 0.092 & 0.095 & 0.114 & 0.028 & 0.028 & 0.043 & 30.000 & 30.000 & 30.000 & 1.000 & 1.000 & 1.000 & 1.000 & 1.000 & 1.000 & 0.051 & 0.052 & 0.042 \\
					Oracle & 0.086 & 0.087 & 0.106 & 0.024 & 0.024 & 0.035 & 10.000 & 10.000 & 10.000 & 1.000 & 1.000 & 1.000 & 0.000 & 0.000 & 0.000 & 0.068 & 0.061 & 0.051 \\
					\hline
			\end{tabular} }
			\captionof*{table}{\scriptsize{Note: All values are based on Monte Carlo simulations with 1000 runs and 1000 repeated subsample steps ($C=1000$). Rejection rates are based on t-tests with heteroscedasticity consistent standard errors (HC3, see \citet{MacKinnon1985}). The remaining measures are means over the 1000 replication runs. PL Stab and DB Stab stand for post-Lasso and double selection with stability selection and the corresponding minum thresholds $\pi_{\mathit{min}}$. Oracle is similar to Fixed Effects All but using only true influencing covariates. \textit{inf} indictates the strength of influential observations and is reported for each measure, while the form of influence (active/inactive set and response) is depicted in the rows.}}
		\end{table}
\end{landscape} }

\begin{table}[]
	\caption{Estimates of the Causal Effect of Tuition Fees $\beta_{(0)}$ for $\theta^*$ in Different Time Frames with HC3 Standard Errors}
	\label{Tab:Fees_Overview_Robust}
	\resizebox{\textwidth}{!}{%
		\begin{tabular}{@{\extracolsep{3pt}}lccccc}
			\hline\hline
			&                                        \multicolumn{3}{c}{Data sets}                                         &  & No. of Variables \\ \cline{2-4}\cline{5-6}
			Tuition Fees                                       &                All                &                Fees                &                Small                &  &  All/Fees/Small  \\ \hline
			&                                   &                                    &                                     &  &                  \\
			\textit{min MSD with} $\theta^*$:                  &         $\mathit{0.9927}$          &          $\mathit{0.9924}$          &          $\mathit{0.9934}$           &  &                  \\
			All Controls                                       &  $\underset{(1.229)}{-1.267 }^{}$  &  $\underset{(1.454)}{-1.952}^{}$   &                  -                  &  &     19/19/-      \\
			Post-Lasso Stability                               &  $\underset{(1.354)}{-2.538}^{}$  &  $\underset{(1.358)}{-2.599}^{}$  &  $\underset{(1.698)}{-6.345}^{**}$  &  &      4/4/3       \\
			Double Selection Stability                         & $\underset{(1.372)}{-3.996}^{**}$ & $\underset{(1.388)}{-3.180 }^{*}$ & $\underset{(3.269)}{-16.468}^{***}$ &  &      7/6/7       \\
			&                                   &                                    &                                     &  &                  \\
			\textit{min MAD with} $\theta^*$:                  &         $\mathit{0.9927}$          &          $\mathit{0.9926}$          &          $\mathit{0.9945}$           &  &                  \\
			All Controls                                       &  $\underset{(1.229)}{-1.267}^{}$  &  $\underset{(1.456)}{-1.941}^{}$   &                  -                  &  &     19/19/-      \\
			Post-Lasso Stability                               &  $\underset{(1.354)}{-2.538}^{}$  &  $\underset{(1.363)}{-2.599}^{}$  & $\underset{(1.745)}{-6.126}^{**}$  &  &      4/4/3       \\
			Double Selection Stability                         & $\underset{(1.372)}{-3.996}^{**}$ & $\underset{(1.392)}{-3.185 }^{*}$ & $\underset{(3.349)}{-17.133}^{***}$  &  &      7/6/7       \\
			&                                   &                                    &                                     &  &                  \\
			$y_{i,t}^{\mathit{extra}}$ with $\pi_{1}$/$\pi_2$: & $\mathit{0.999} / \mathit{0.9} $  &  $\mathit{0.9} / \mathit{0.9} $   &   $\mathit{0.85} / \mathit{0.91} $   &  &                  \\
			All Controls                                       &  $\underset{(0.922)}{-1.722}^{}$  &  $\underset{(1.087)}{-2.213}^{*}$   &                  -                  &  &     19/19/-      \\
			Post-Lasso Stability                               & $\underset{(1.369)}{-3.349}^{*}$  &  $\underset{(0.942)}{-2.234}^{*}$  &   $\underset{(1.519)}{-11.570}^{***}$    &  &  3/9/2      \\
			Double Selection Stability                         & $\underset{(1.151)}{-3.920}^{***}$  & $\underset{(1.000)}{-2.198}^{*}$  &  $\underset{(4.415)}{-15.021}^{**}$  &  &      6/10/6      \\ \hline\hline
		\end{tabular}%
	}
	\caption*{\scriptsize{Note: Response values are scaled to a percentage level. Standard errors in parentheses are heteroscedasticity consistent (HC3, see \citet{MacKinnon1985}). $^{*}$p$<$0.05; $^{**}$p$<$0.01; $^{***}$p$<$0.001 indicate p-values from a t-test on significance from zero. $\theta^*$ is chosen according to minimum mean squared deviation (MSD) and minimum mean absolute deviation (MAD).}}
\end{table}

\end{document}